\pdfoutput=1

\documentclass[12pt,oneside,letterpaper]{book}

\usepackage[]{uahdis,chngpage,units,bm,hhline,rotating,verbatim,amsfonts,amssymb,xspace, textcomp}
\usepackage[]{indentfirst, booktabs, enumerate, gensymb}
\usepackage{color,soul}
\usepackage[round, sort]{natbib}
\usepackage{url}
\usepackage{multirow}
\usepackage{listings}
\usepackage{etoolbox}
\apptocmd{\sloppy}{\hbadness 10000\relax}{}{}

\author{Zachary Donald Robinson}

\title{New Probabilistic Model for Episode Integrated Fluences of Protons using Episodes from 1973-2013}

\date{2015}

\uahdepartment{Physics}

\uahadvisor{James H. Adams, Jr.}

\uahmema{Nikolai Pogorelov}

\uahmemb{Gang Li}

\uahmemc{Qiang Hu}

\uahdeptchair{James A. Miller}

\uahcollege{Science}

\uahcolldean{Sundar Christopher}

\uahdegree{Master of Science}

\uahprogram{Physics}

\uahdegreeshort{master's}

\uahdoctype{thesis}

\uahgraddean{David Berkowitz}

\begin{document}


\frontmatter

\pagestyle{plain}

\maketitle

\copyrightpage

\approvalpage


\chapter*{Abstract}
\makeabstract


A new probabilistic model for protons has been created using episode integrated fluences. This model will allow the user to choose a mission start date, mission duration, and confidence level to construct a mission-specific, bounding case spectrum for proton fluences at a distance of 1 $AU$ from the sun. A new database of episode integrated fluences will be created for this model. The database will contain 29 channels that span the energy range 0.88-485 $MeV$. This database will cover the period from November 1, 1973 to December 31, 2013, making it the largest database on solar activity.


\abstractsig


\chapter*{Acknowledgments}

The work described in this thesis would not have been possible without the assistance of a number of people who deserve special mention. First, I would like to thank Dr. James H. Adams, Jr. for working with me the last few years. I appreciate all the guidance and help that you have provided me while working on this thesis. Second, I would like to thank the other members of my committee for your very helpful comments and suggestions. Finally, I would like to thank my family and friends who have encouraged (and nagged) me while writing this thesis. Thank you all very much.

\addtocontents{toc}{\protect\hfill{PAGE}\par}
\tableofcontents

\listoffigures

\listoftables


\addchapheadtotoc



\clearpage

\pagestyle{myheadings} \markright{}

\mainmatter

\chapter{Introduction}
\label{intro}
Space weather is an important area of research during the current space age of humanity. Spacecraft designers and mission planners are responsible for delivering the spacecraft to space, but also making sure that it operates reliably during its mission. Every mission has its own set of hurdles and challenges that it will have to overcome. But every mission has to deal with the space radiation environment. This environment is comprised of energetic charged particles of solar or galactic origin that are moving through space. If these particles are energetic enough, they can penetrate into the spacecraft and, if they are sufficiently ionizing, they can destroy the electronics and pose a hazard to space crews. For these reasons, spacecraft designers and mission planners carefully evaluate the radiation shielding provided by the spacecraft and add shielding as needed to protect the electronics and crew inside. One of the main issues is how much shielding is necessary to protect the electronics and crew throughout the entire mission. Should the spacecraft be designed to withstand the worse possible solar energetic particle (SEP) events that have ever been observed (e.g.\ the Carrington Event of 1859 or the October 1989 event)? Is it possible to have less shielding and only protect the payload against smaller SEP events because a larger event is so unlikely to occur during the mission? These are just some of the questions that spacecraft designers and mission planners must ask themselves during the designing and planning process. 

Over the last few decades, tools have been developed to help mission planners and designers attempt to answer these questions. Since protons are the most abundant particle found in SEP events, most tools set out to try to predict the proton component of the radiation environment. These tools, or models, attempted to predict the occurrence of the SEP events and their sizes during a given time period. These models are capable of producing a prediction of the proton environment for a given time period. These predictions take the forms of cumulative fluence \citep{King74, Xapsos99b}, event-integrated fluences \citep{Feynman90, Xapsos99a}, or peak flux \citep{Xapsos98, Nymmik99}. 

One of the few similarities between all the models is their focus on SEP events. Events are caused by the explosive release of energy stored in the magnetic fields in the solar photosphere. These explosive outbursts cause both solar flares and coronal mass ejections (CME). The flares accelerate particles and produce intense electromagnetic radiation, including X-rays. Coronal mass ejections are clouds of plasma that travel outwards from the sun at very high speeds so that they exceed the speed of sound when they reach the chromosphere and begin to drive a shock which accelerates particles to high energies. These particles escaping from both shock and flares travel along the magnetic field lines extending out from the sun. These traveling particles form the majority of the ionized radiation found around Earth during an event. These events are said to begin at Earth when a particle flux is detected above the normal particle background level, and end once the flux returns to background. However, more times than not, the particles from a second event reach Earth before the flux from the first event reaches background. This phenomenon is referred to as an episode.

A new model will be presented in this thesis that will allow mission planners and spacecraft designers to construct a mission-specific, bounding case spectrum of proton radiation at a distance of 1 $AU$ from the sun. This model will allow the user to specify mission start date and mission duration and then construct the bounding case spectrum at a user-specified confidence level. The model will be able to construct a spectrum for a mission occurring during the time period 1953-2052. The model will use extreme value theory to create the bounding case spectrum. Episode fluence will be used in this model instead of the traditional event fluence. The model's use of episode fluences differentiates this new model from all the previous ones. Since the particle flux that a spacecraft will experience at a given moment does not always come from just one event, it is better to focus on episodes of events rather than single events. A database of episode fluences spanning the energy range 0.88-485 $MeV$ will also be presented in this thesis. Since past models have always used event fluences, this new database was constructed for use in the probabilistic model.

In this thesis, the database of episode fluences and the new proton radiation model will be discussed. \chapterref{background} will discuss the background of modern solar proton event models used for predicting proton radiation. \chapterref{data} will describe in detail how the new database of episodes fluences was created. This database is the foundation on which the probabilistic model is based. \chapterref{method} will discuss how the model uses extreme value theory to create the bounding case fluence spectrum for a mission. This chapter will also include sections on how the database of episode fluences was used to find the cumulative distributions and the number of episodes per year. The results of testing the probabilistic model will also be discussed. \chapterref{conclusion} will discuss the real world use for this model and potential areas of future research.

\chapter{Background}
\label{background}
The modern study of solar particle events started around 1956. Before this time, there were only four instances of increased solar cosmic-ray activity measured at the surface of the Earth in the previous 15 years that could be related to specific solar activity. This led to the assumption that these events were quite rare \citep{Webber63}. There was also not much knowledge of the radiation spectrum that these events produced. However, as technology developed and better tools were produced to study these events (balloon flights, satellites, space probes, etc.), scientists were able to describe a more complete picture of these events. The composition of the cosmic-ray particles ejected from the sun were primarily protons that had steep energy spectra \citep{Webber63}. These protons had energies ranging from less than 10 Mev to a few Gev for the larger solar energetic particle events \citep{Webber63}. These particles arrived at Earth shortly after being produced in the region of a solar flare (although today it is understood that most solar energetic particles come from coronal mass ejections, or CMEs \citep{Hundhausen99}). The intensity of these events was also found to exceed the galactic cosmic-ray intensity by thousands of times for a period of a few days \citep{Webber63}. Even though some of the solar events had similar characteristics, most of the time these characteristics differed on an event to event basis.

Eventually, models were built to predict the flux or fluence of protons at Earth coming from these events. One of the earliest models was the King model. The King model gave the probability that any given solar proton fluence level will be exceeded during a space mission that occurred during solar maximum of solar cycle 20 \citep{King74}. This model used fluence level, proton energy threshold, and mission duration as its parameters. The energy range of this model is 10-100 MeV. The calculations used in this model are only based on the data collected from 1966-1972 or roughly just the previous solar maximum. The downsides to this model are that it is only designed for solar cycle 20 and that it is based on a small data set, which is used to compute the probability.

The King model was used as the standard model until roughly 1990. The JPL model \citep{Feynman90} was created to try to improve on the predictive abilities for the fluence of SPEs. The JPL model looked at event-integrated fluences for energies $>10$ $MeV$ and $>30$ $MeV$ \citep{Feynman90}. The JPL model found good ageement with the King model at energies $>30$ $MeV$ but found the fluence to be twice that expected in the King model for $>10$ $MeV$ \citep{Feynman90}. This model also used data from 1956 to 1985, spanning a period of time about three times larger than the King model. The JPL model didn't rely on sunspot number or the distinction between ordinary proton events and anomalously large events that was required in the King model \citep{Feynman90}. \citet{Feynman02} showed that the JPL model was still valid and in agreement with the additional data and the newer models developed.

In 1998 and 1999, Michael Xapsos developed models for the peak flux, event-integrated fluence, and mission-integrated fluences for protons. Both the peak flux and the event-integrated fluence models use the maximum entropy principle \citep{Xapsos98, Xapsos99a}. The peak flux model creates mission specific, worst case, $>10$ $MeV$ solar proton event peak fluxes. This model also predicted that there is an upper limit that the peak flux can attain, which is approximately twice the largest peak flux on record at the time \citep{Xapsos98}. The event-integrated fluence model predicted the worst case event fluence for a user specified time interval. This model also predicted the proton energies in the largest range available at the time, $>1$ $MeV$ to $>300$ $MeV$ \citep{Xapsos99a}. The third model (known as the ESP model) developed by Xapsos during this time period predicted the cumulative proton fluence for a user specified mission duration \citep{Xapsos99b}. For the energy ranges $>1$ $MeV$ to $>100$ $MeV$, there was enough data to do a statistical model for these energy ranges at this time. This wasn't the case for the ranges $>100$ $MeV$ to $>300$ $MeV$ so an empirical approach was taken. The data used in all three of these models spanned roughly the same time period, 1963 to 1996. This corresponds to solar cycles 20, 21, and 22. \citet{Xapsos04} updated the ESP model with the more recent data. 

There is a model by Riho Nymmik that was proposed as the International Standard \citep{Nymmik99}. This model creates a probability for $\geq10$ $MeV/nucleon$ fluences and peak fluxes. This model would not only be able to calculate the probabilities for protons, but also for Z=2 to Z=28 ions \citep{Nymmik99}. 

In 2005, \citeauthor{Rosenqvist05} proposed an update to the JPL model. The main update to the JPL model was to replace the Monte Carlo method in it with an analytic solution. \citeauthor{Rosenqvist05} planned to make the JPL model into one that is a fully reproducible computer-based procedure so that it could be easily applied to any new data set and not just data in the format needed in the JPL model \citep{Rosenqvist05}. The model used a dataset that consisted of events from 1974 to 2003. \citeauthor{Rosenqvist05} also wanted to created a data set that was general enough so that it could be used to check other models, for example \citeauthor{King74}, \citeauthor{Nymmik99}, and \citeauthor{Xapsos98} \citep{Rosenqvist05}.

\citet{Jun07} used a Monte Carlo approach to estimate worse case mission integrated proton fluence. This paper focused on the statistical distributions of event fluences, event duartions, and time intervals between adjacent events \citep{Jun07}. The event fluences could be fit to a log-normal distribution while the event durations and time interval between events can be represented by the Poisson Distribution \citep{Jun07}. 

\citet{Jiggens12} developed a model to calculate the peak flux of SPEs by using the L\'{e}vy distribution. The authors used the L\'{e}vy distribution instead of the Poisson distribution because the L\'{e}vy distribution is time-dependent and they wanted to create a model that did not make the time-independent assumptions needed for a Poisson process. 

There are other models that were created but are not used as frequently or did not make major imporvements to the existing models. \citet{Burrell71} calculated tissue doses in rads at the center of an aluminum spherical shell. The SOLPRO model \citep{Stassinopoulos74} was a model that was based off of a solar cycle that was dominated by 1 large SPE. GOST 2545.134-86 is a model that uses a log-normal distribution to describe the occurrence of events during the solar maximum \citep{GOST86}. CREME86 offered a lot of different choices for the SEP environments \citep{Adams81, Adams85}. CREME96 is an updated version of CREME86 that contains a model for the worst day and worst week SPE environments based on the October 1989 event \citep{Tylka97}. \citet{Xapsos00a} presented a model that predicted cumulative solar proton event fluences for a space mission. \citet{Gerontidou02} looked at the frequency distributions of peak intensities of solar proton events for 1976-1999. \citet{Xapsos08} created a long term solar energetic particle event environment model. \citet{Kim09} used the occurrence of large SEP events to predict doses in typical blood-forming organs. \citet{Schwadron10} proposed the groundwork for a model that will predict how the radiation environment evolves as a function of radial distance from the sun. This paper also identifies areas that still need further research before the model can be implemented. \citet{Laurenza12} studied the time evolution of SPE spectrum by applying Shannon differential entropy. \citet*{Usoskin12} tried to find a maximum fluence of protons in solar particle events on a time scale of tens of millennia. \citet*{Kovaltsov14} looked at lunar rocks to try to predict the maximum proton fluence in a year over a period of a million years.    

These models are all good and useful but they all use events. The model proposed in this thesis will create the bounding case spectrum by using episodes instead of events. This model will also use the largest database of proton fluence measurements of any model.

\chapter{Episode Database}
\label{data}
The following sections will describe the steps taken to process the data from the two satellites used to create the database of episode fluences. The chapter begins by describing the two satellites in detail and discussing the steps used to identify episodes in the data. Next, the formation of a seamless database from two different satellites will be discussed. Finally, there will be a section describing the typical characteristics of episodes.

\section{Data Processing}
\label{dataprocessingsection}

The method used for defining episodes creates a database in which episodes are so weakly correlated that they can be treated as statistically independent. Actually, these episodes are not completely statistically independent. However, if it can be assumed that an active region on the sun does not contribute to more than one episode, then the occurrence of episodes will be approximately independent.  Although active regions have been found in the database that contribute events to more than one episode, they are rare ($<10\%$). In what follows, it will be assumed that the episodes are statistically independent and time-independent Poisson statistics will be used to describe them. 

The database that was created for this model uses the instruments on two different satellites. The earlier data comes from the Goddard Medium Energy (GME) experiment on IMP-8. The Energetic Particle Sensors (EPS) instrument flown on the GOES series provides the later data used in this database. The following two sections will describe the two instruments and the steps taken to process the data.

\subsection{GME Data}
\label{GMEdata}

The GME data used here was measured by the Goddard Medium Energy Experiment on the Interplanetary Monitoring Platform-8 (IMP-8) and is described by \citet{McGuire96}. IMP-8 was launched October 26, 1973. Its orbit is approximately circular at 35 Earth radii so its instrumentation was well positioned to measure interplanetary solar particle fluxes. The GME data discussed here is available with 30 minute resolution over the time period from November 1, 1973 to October 31, 2001 on the Coordinated Data Analysis Web. Although contact with IMP-8 continued for a few years, there are large data gaps after October 31, 2001 due to periods of non-communication with the satellite. The GME data processing was completed by Mike Xapsos and Craig Stauffer \citep{Xapsos05}. The following paragraphs summarize the steps used in \citet{Xapsos05} to process the GME data.

GME has a total of 30 channels that measure proton flux data. These 30 channels cover most of the range from 0.88-485 MeV. There is a 6 MeV gap (from 81 MeV to 87 MeV) between channels 22 and 23. Interpolation was used to extend each channel to 83.95 MeV in order to create a continuous data set. Channel 15 was excluded from the data set since most of this channel's energy range is covered by channel 16. This created a second data gap from 18.7 MeV to 19.8 MeV in between channels 14 and 16. Both channels were extended to 19.24 MeV using interpolation to fill this gap. The channel number was also reduced by one beginning with channel 16 to have sequentially numbered channels. The energy bin boundaries for the 29 GME energy channels used are given in \tableref{GMEchannels}.
\begin{table}
	\begin{center}
		\caption{A list of the GME channels and their widths that were used in the GME data portion of the episode database.}
		\label{GMEchannels}
		\begin{tabular}{|c|c|c|}
			\hline
			GME & Lower  & Upper \\ 
			Channel & Energy (MeV) & Energy (MeV) \\ \hline
			1 & 0.88 & 1.15 \\ \hline
			2 & 1.15 & 1.43 \\ \hline
			3 & 1.43 & 1.79 \\ \hline
			4 & 1.79 & 2.27 \\ \hline
			5 & 2.27 & 3.03 \\ \hline
			6 & 3.03 & 4.2 \\ \hline
			7 & 4.2 & 4.94 \\ \hline
			8 & 4.94 & 5.96 \\ \hline
			9 & 5.96 & 7.25 \\ \hline
			10 & 7.25 & 8.65 \\ \hline
			11 & 8.65 & 11.1 \\ \hline
			12 & 11.1 & 13.6 \\ \hline
			13 & 13.6 & 16.1 \\ \hline
			14 & 16.1 & 19.24 \\ \hline
			15 & 19.24 & 24.2 \\ \hline
			16 & 24.2 & 28.7 \\ \hline
			17 & 28.7 & 35.2 \\ \hline
			18 & 35.2 & 42.9 \\ \hline
			19 & 42.9 & 51 \\ \hline
			20 & 51 & 63.2 \\ \hline
			21 & 63.2 & 84 \\ \hline
			22 & 84 & 92.5 \\ \hline
			23 & 92.5 & 107 \\ \hline
			24 & 107 & 121 \\ \hline
			25 & 121 & 154 \\ \hline
			26 & 154 & 178 \\ \hline
			27 & 178 & 230 \\ \hline
			28 & 230 & 327 \\ \hline
			29 & 327 & 485 \\ \hline
		\end{tabular}
	\end{center}
\end{table}

The GME data in this work have been assumed to provide absolute values of proton flux. This is based on the appropriateness of its orbit for measuring interplanetary proton fluxes, the energy resolution and the well-defined detector geometry factors of the experiment package. The GME instrument has narrower energy channels which allow for more information to be gathered on the particles present during an episode. \citet{Rosenqvist05} determined that the most reliable instruments to use were IMP-8/GME, GOES-7, and GOES-8. The disadvantage of GME is the large number of data gaps in the record. The GOES satellites have much better coverage. 

The GME data set contains some gaps, which may be caused by incomplete data recovery by the tracking network, telemetry errors, or instrument saturation. These gaps were identified and filled as follows. Small data gaps can be filled reliably by interpolation using the good data preceding and following the gap.  This was done using either a linear or logarithmic interpolation depending on which was most consistent with the time profile of the event at the point where the gap occurs. Larger data gaps exist, for example, when the GME instrument saturated. To fill these gaps, the time profile had to be obtained from an alternate data source. For gaps occurring before 1986, data was used from the Charged Particle Measurement Experiment on IMP-8. From 1986 until the data record ends in 2001, data was used from the GOES satellites to fill gaps. The data from the alternate sources (used to fill broad data gaps) was scaled to match the GME flux preceding and following the gap. 

In addition, there is a background cosmic ray particle flux that is incorporated in the raw GME data. To subtract out this background, a quiet period of time spanning anywhere from a few days to two weeks was looked for in each year. The flux during this time was then averaged to find the background value for a GME channel. This was done for all 29 channels for each year in the GME data. The periods of time used for the background were permitted to be different for each channel in a given year. Periods of time were avoided if another channel had increased flux during the same interval. This was done to avoid introducing any residual background effects into the episode fluences. 

Now, with a complete gap-free and background-free data set, a search was made for episodes of elevated proton flux. These episodes were identified when the peak flux exceeded either 4 $cm^{2}s^{-1}sr^{-1}MeV^{-1}$ in the 1.15 to 1.43 $MeV$ energy channel or 0.001 $cm^{2}s^{-1}sr^{-1}MeV^{-1}$ in the 42.9 to 51.0 $MeV$ energy channel. The trigger threshold for the high-energy bin was included to avoid missing small episodes that consist of hard energy spectra events. Such episodes can be submerged in the background at low energies, only appearing above background in the higher energy channels. Only one episode in the data set was found that exceeded the high energy threshold without triggering the low energy threshold. 

Whenever possible the onset and end times of episodes were determined from the time at which the flux first exceeded background to the time at which it returned to background. Sometimes the flux fell below the threshold and then rose above it before falling to background. When this occurred, the second rise above threshold was interpreted as the start of a second independent episode. In these cases, the two episodes were separated by the local minimum in the flux-time profile. If the flux showed successive rises and falls but remained above the threshold, only one episode was identified. To calculate the episode-integrated fluences, the 30-minute averaged fluxes are multiplied by 1800 to convert them to fluence. The fluences for all the 30-minute intervals of an episode were summed up to get the total fluence of the episode. This procedure was repeated for each energy channel.

\subsection{GOES Data}

The GOES data comes from the Energetic Particle Sensors (EPS) instrument. This instrument, along with the High Energy Proton and Alpha Detector (HEPAD) instrument, make up the Space Environment Monitor (SEM) package that is flown onboard the Geostationary Operational Environmental Satellite (GOES). There are multiple satellites in the GOES series, with the first one launched in May of 1974. All of the GOES satellites orbit the Earth in geostationary orbit, approximately 35,800 km above Earth's surface \citep{Onsager96}. Each satellite in the series was given a letter before it was launched and a number once the satellite is in orbit. For example, GOES-I was renamed to GOES-8 once in orbit \citep{namechange}. The GOES satellites were used from November 1, 2001 to December 31, 2013. NOAA's recommendations \citep[see][]{primary} for the primary satellite for proton measurements were followed. This led to GOES-8 being used from November 1, 2001 to May 31, 2003, GOES-11  from June 1, 2001 to April 30, 2010, and GOES-13  from May 1, 2010 to December 31, 2013. The energy channels for the EPS instrument can be found in \tableref{GOESchannels}. A complete description of these instruments can be found in \citet{Onsager96}. 
\begin{table}
	\begin{center}
		\caption{A list of the GOES channels and their widths that were used in the GOES data portion of the episode database \citep{Onsager96}.}
		\label{GOESchannels}
		\begin{tabular}{|c|c|c|}
			\hline
			GOES & Lower  & Upper \\ 
			Channel & Energy (MeV) & Energy (MeV) \\ \hline
			P1 & 0.7 & 4.2 \\ \hline
			P2 & 4.2 & 8.7 \\ \hline
			P3 & 8.7 & 14.5 \\ \hline
			P4 & 15 & 40 \\ \hline
			P5 & 38 & 82 \\ \hline
			P6 & 84 & 200 \\ \hline
			P7 & 110 & 900 \\ \hline
		\end{tabular}
	\end{center}
\end{table}

The data in GOES channel P1 is usually contaminated by particles of magnetospheric origin. These magnetospheric particles caused fluctuations in channel P1 that were seen daily over many months in all three GOES satellites used in this analysis. This time-varying background made it difficult to identify episodes in this channel. Generally, the solar energetic particle flux in P1 only exceeded the magnetospheric background during relatively large solar energetic particle events. Channel P1 was excluded from the dataset due to the difficulty in identifying episodes that were caused by this contamination.

In this dataset, the raw GOES data refers to the data found on NOAA's website that had the Zwickl process preformed on it \citep[see][]{GOESdata}. The Zwickl process is explained on NOAA's website \citep[see][]{Zwickl}. The raw data from GOES also had some bad data scattered through it. For GOES-8 and GOES-11, these are marked by the value 32700 while GOES-13 used -99999 to mark bad data. These bad data have to be corrected through interpolation or substitution. If there were only one or two successive bad 5-minute average fluxes, the bad data was replaced by linearly interpolating between the preceding and following good data. When three or more bad 5-minute average fluxes appear in succession, the value 9.0E6 was inserted as a temporary bad-data marker. These larger gaps were not replace with data from other sources at this point because they were only filled if the gap occurred during an episode. The raw data files that were corrected using this procedure were relabeled as cleaned data files.

Even though the raw data had the Zwickl process preformed on it, the cleaned data still include residual background from cosmic rays and particles of magnetospheric origin.  These backgrounds must be removed. The cleaned 5-minute averaged fluxes were combined to obtain daily flux averages and were then plotted for each year. In each annual plot, a search was made for a long period having the lowest fluxes in each energy channel. The daily-averaged fluxes in each channel were averaged over these periods to obtain estimates of the residual backgrounds for each energy channel. An objective of this search was to include the maximum numbers of days into these averages in order to reduce the effects of daily fluctuations. These background fluxes were subtracted from the cleaned 5-minute-averaged flux data. When the 5-minute averaged flux was less than the background level, the background-subtracted flux was set to zero. 

It was decided to use 30-minute averages of the cleaned and background-subtracted fluxes to identify episodes because their higher statistical precision outweighed the resulting loss in temporal resolution. This choice was confirmed by comparing GME and GOES measurements during the period from January 1, 2000 to October 31, 2001 when measurements from both satellites were available. Both 5-minute and 30-minute averages were used to determine onset and end times for episodes of elevated proton flux. It was concluded that the 30-minute averages gave start and stop times for the episodes that were in closer agreement with those determined from GME. All the cleaned and background-subtracted 5-minute data for each month was then converted in 30-minute flux averages and stored in separate files. 

Episodes of elevated proton flux were identified by graphing the data from channels P2 through P7 for each month from November 1, 2001 to the end of 2013. These graphs were used to find the onset and end times for each episode. An episode was recognized when the flux was seen above the residual background for an extended period. The onset of each episode was identified as the point at which the flux first rose above the residual background level. The end was identified as the point at which the flux returns to the residual background. Onset and end times were identified in channels P2 through P7. The onset and end times in channel P2 were used to define the temporal extent of individual episodes. Occasionally in higher energy channels, the onset would occur earlier than in the lower energy channels, with the flux dropping to background earlier as well. It also sometimes happened that the flux in a higher energy channel dropped to background and then rose again. In these cases, a second onset and end time was determined for the channel but both periods of elevated flux were defined as part of the same episode. After the onset and end times were identified in each energy channel, the fluxes in each channel were converted to fluences as described in the previous section (\sectionref{GMEdata}).

The episode-integrated fluences for each episode were checked for extremely high values, which indicated the presence of one or more data gaps in the episode that had been filled with flux values of 9.0E6 during the cleaning process. The monthly plots of the 30-minute averaged flux for the episode were examined for the channel having the high fluence to find the data gap or gaps during the episode. These data gaps were filled in the clean 5-miunute data using data from one of the secondary GOES satellites recommended by NOAA. In those cases where the secondary satellites had a data gap at the same time, interpolation of the primary satellite's data was used. It should be noted that this form of interpolation was never needed for a period longer than two hours.

\section{Creating a Seamless Database}

Now that the data has been cleaned and episodes identified, the next step was to create a seamless dataset. There are two hurdles that have to be overcome to achieve this seamless dataset. The first is the instrumental differences, especially those between GME and GOES. The second is that GME and GOES have different energy bins or channels.

\subsection{Normalization}

In order to create a single internally consistent data set, it is necessary to normalize the measurements. GME was chosen as the benchmark instrument based on investigations of various instruments that have been reported in the literature \citep{Rosenqvist05}. The GME energy channels are more finely spaced and would provide a better bounding case spectrum for the probabilistic model. After GME coverage ended in October of 2001, GOES-8 was chosen to extend the observations since there is a general agreement between GOES and GME to within a factor of 2 \citep{Smart99}. GOES-8 made measurements during the maximum of solar cycle 23, providing strong overlap with GME measurements.

To normalize GOES-8 to GME, fluences measured by GME were integrated  between the lower and upper energy boundaries of the GOES channels for the period January 1, 2000 to October 31, 2001. This was done for each GOES channel. Fluence measurements made over the same time intervals by GOES-8 were multiplied by the GOES energy channel widths to obtain measurements which could be compared with the GME fluence integrated over the GOES energy channels.

Next, for every episode during this time, the flux in each channel was examined to determine whether the majority of the episode fluence came from the episode itself and not background. The goal here was to determine whether the average episode flux was 10 times greater than the background flux. The episodes that fulfilled this requirement were marked as `good'.  For the larger episodes, this was an easy classification because the episode flux dominated for the entire length of the episode. The smaller episodes had to be carefully examined by eye to estimate if the average episode flux was 10 times greater than the average background flux for the entire episode. The episodes were examined for peaks of high flux. A high peak can dominate an episode that is short even if for most of the episode the flux is close to background. The deciding factors for these cases were how much higher the peak is compared to the background, how long the peak persists at the high flux level, and how long the episode is close to background. After every channel of every episode for both satellites was examined, a GME to GOES-8 fluence ratio was created for all the channels where both satellites were marked as having the average episode flux at least 10 times greater than the average background flux. With these GME to GOES-8 ratios, the scaling factor for each channel could be found by taking the average of the ratios for those episodes that was classified as `good' in that energy channel. These scaling factors were used to normalize the GOES-8 fluences to GME fluences.

For the later GOES data, GOES-11 and GOES-13 also had to be normalized to GME. The normalization factors between all the GOES satellites have been reported by \citet{Rodriguez14}. Since there was no scaling factor for GOES-11 or GOES-13 to GOES-8, intermediate satellites were used. GOES-13 was normalized to GOES-8 via GOES-10 while GOES-11 was normalized to GOES-8 via GOES-10 and GOES-13. The GOES-8 normalization to GME was then used to normalize GOES-11 and GOES-13 to GME. The normalization factors for each GOES satellite to GME is presented in \tableref{normalization}.
\begin{table}
	\begin{center}
		\caption{The GME normalization factors for each of the three GOES satellites.}
		\label{normalization}
		\begin{tabular}{c|c|c|c|}
			\cline{2-4}
			& \multicolumn{3}{c|}{GOES Satellites} \\ \hline
			\multicolumn{1}{|c|}{Channel} & GOES-8 & GOES-11 & GOES-13 \\ \hline
			\multicolumn{1}{|c|}{P2} & 1.287 & 1.093 & 1.082 \\ \hline
			\multicolumn{1}{|c|}{P3} & 0.857 & 0.900 & 1.057 \\ \hline
			\multicolumn{1}{|c|}{P4} & 0.567 & 0.664 & 0.511 \\ \hline
			\multicolumn{1}{|c|}{P5} & 0.795 & 0.801 & 0.863 \\ \hline
			\multicolumn{1}{|c|}{P6} & 1.498 & 1.413 & 1.323 \\ \hline
			\multicolumn{1}{|c|}{P7} & 0.230 & 0.248 & 0.235 \\ \hline
		\end{tabular}
	\end{center}
\end{table}

\subsection{Fitting the GOES Spectra}
\label{fittingSpectra}

\begin{figure}
	\begin{center}
		\includegraphics[scale=0.75]{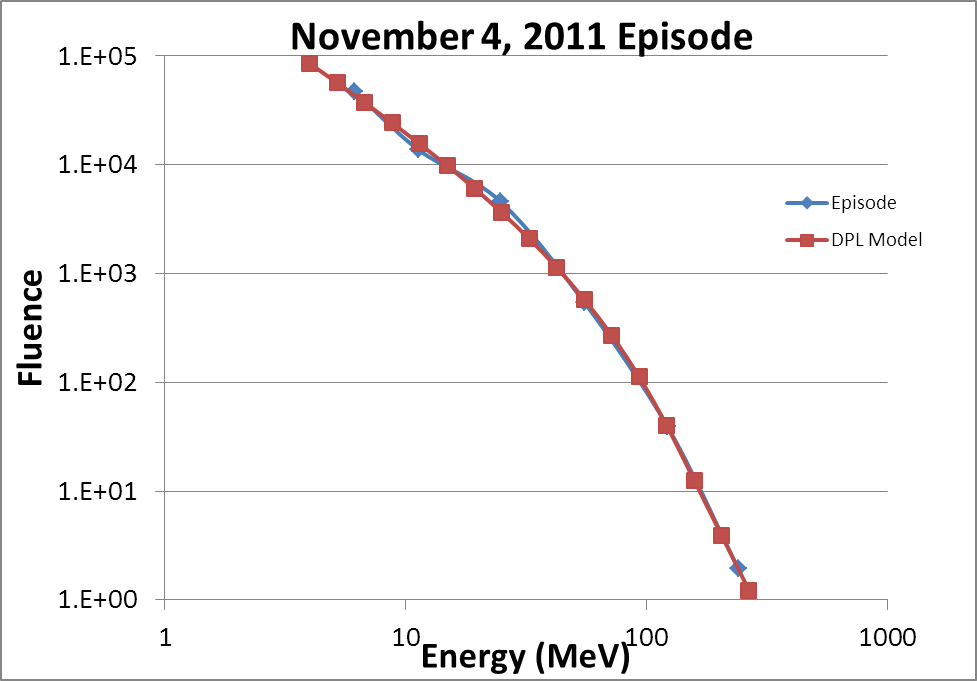}
		\caption{The Band Function, or the Double Power Law, is fitted to the November 4, 2011 episode. The $\chi^2$ value for this fit was 0.09433.}
		\label{Band}
	\end{center}
\end{figure}

\begin{figure}
	\begin{center}
		\includegraphics[scale=0.75]{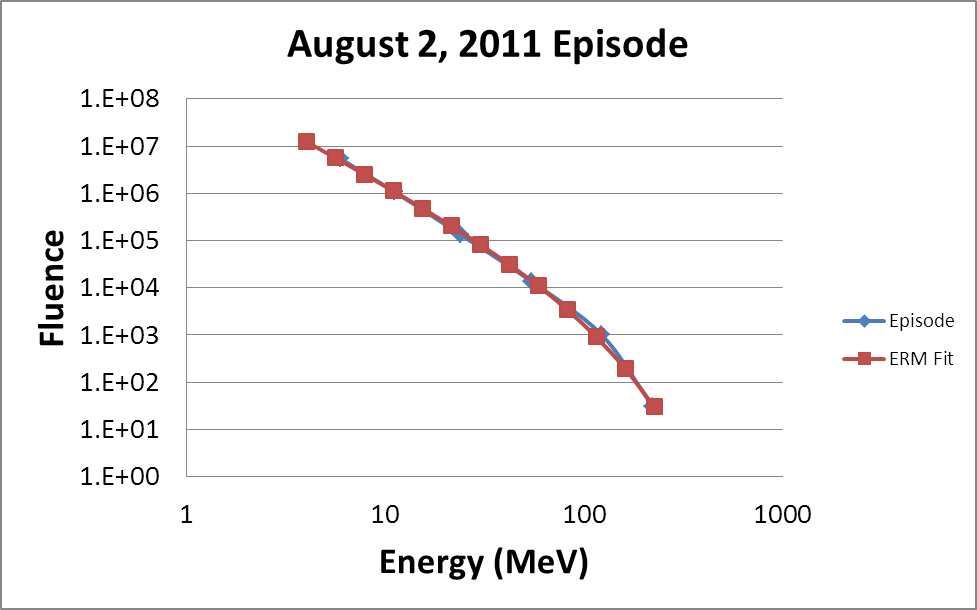}
		\caption{The Ellision-Ramaty model is fitted to the August 2, 2011 episode. The $\chi^2$ value for this fit was 1.191.}
		\label{ERM}
	\end{center}
\end{figure}

\begin{figure}
	\begin{center}
		\includegraphics[scale=0.75]{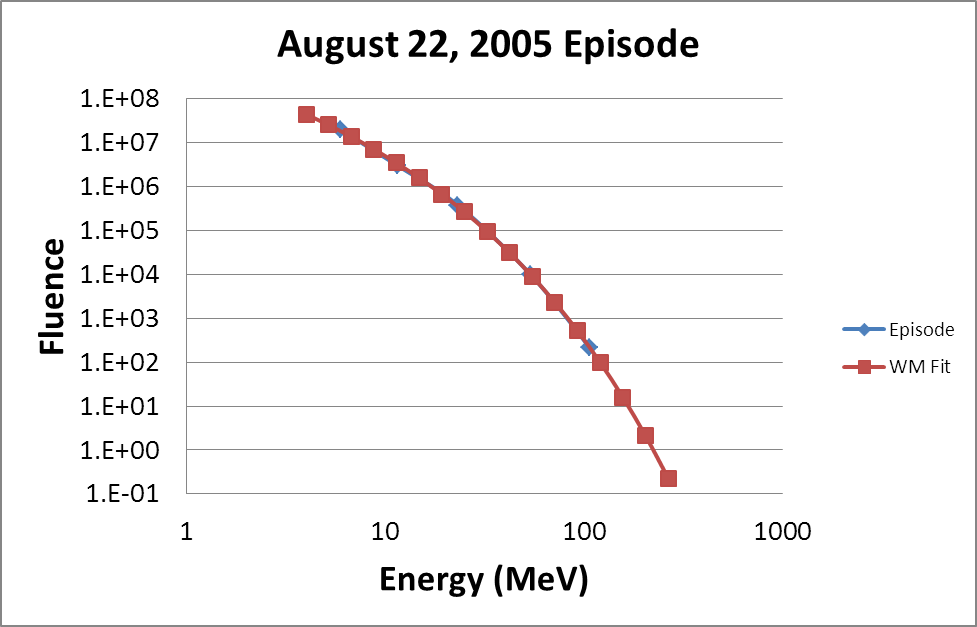}
		\caption{The Weibull model is fitted to the August 22, 2005 episode. The $\chi^2$ value for this fit was 0.1744.}
		\label{WM}
	\end{center}
\end{figure}

Now that the instrumental differences have been nullified, the final step to creating a seamless data set is to redistribute the data so that both satellites use the same energy channels. This was done by fitting the GOES fluence spectra to spectral representations found in the literature. The differential fluence spectra were fit using four trial spectral models: the Band Function \citep{Band93}, the Ellision-Ramaty model \citep{Ellison85}, the Weibull model \citep{Xapsos00b}, and a power law in energy, $f(E)=AE^{-\gamma}$. Examples of the Band Function, Ellision-Ramaty model, and the Weibull model fits are plotted in \figureref{Band}, \figureref{ERM}, and \figureref{WM}, respectively.

Because of the width of the GOES energy channels, it was necessary to carefully choose the median energy value, $\overline{E}$, for each channel, i.e. a value such that the energies of half the events in the channel fell below and half above this energy value, i.e. a value such that
\begin{equation} \label{midenergy}
\int_{E_{0}}^{\overline{E}} f(E) dE = \int_{\overline{E}}^{E_{1}} f(E) dE,
\end{equation}
where $E_{0}$ and $E_{1}$ are the lower and upper energy bounds of a GOES energy channel. The value of $\overline{E}$ was found by following two recursive procedures. In the first, $f(E)$ was assumed to be a power law. To obtain an initial estimate of $\gamma$, the midpoint energy of each channel was used as the median energy and power law fits between successive energy channels were used to estimate $\gamma$ for each channel pair. The $\gamma$ values thus obtained were interpolated to obtain an estimate of the correct $\gamma$ to use for each energy channel. Equation \eqref{midenergy} was then solved to obtain a refined estimate of $\overline{E}$. Using these refined median energies for each channel, the values of $\gamma$ were recomputed. It took only a few iterations for the values of $\overline{E}$ to converge. The second iterative procedure was to fit the spectrum with the four trial spectral models discussed above in order to find the one that gives the lowest $\chi^{2}$. The model with the lowest $\chi^{2}$ was used in Equation \eqref{midenergy} to refine the estimates of $\overline{E}$ for each energy channel.  The spectrum was then re-fit with the four trial models using the refined values of $\overline{E}$. This second iterative procedure was also found to converge rapidly.

The fitting process was considered successful if the reduced Chi Squared was less than 1.5. When the best fit gave a reduced Chi Square value of 1.5 or higher, the onset and end times for the episode were checked and corrected, if necessary, to remove any residual background. In a few instances, the best-fit Chi Square value remained too high. In these cases, it was found that the episode was dominated by two SEP events, one with a soft energy spectrum and one with a hard spectrum. The soft spectrum dominated at low energies while the hard spectrum dominated at high energies. In these cases, it was necessary to fit the spectrum with a combination of a power law in energy and one of the other fitting trial functions in order to obtain a reduced $\chi^{2} < 1.5$.

There were also some small episodes that only had flux present in channel P2 on GOES. These episodes had to be excluded from the data set since there was no reliable method to accurately fit their fluence spectra. 

The resulting spectral fits for the episodes measured with GOES were used to distribute the GOES fluence measurements into the 29 GME energy channels. The midpoint energy for each GME channel was taken to be the median value since these energy channels are so narrow. These energies were used with the chosen spectral representations to distribute the GOES fluence spectra into the GME channels for all the episodes measured using GOES. The resulting database can be found online \citep[see][]{database}. It consists of the onset and end times together with differential energy spectra of the episode-integrated proton fluences.

\section{Episode Characteristics}

In this section, the characteristics of the episodes with elevated proton fluxes are discussed. The characteristics of these episodes include duration, the number of peaks, steepness of the flux increases and decreases and the identification of peaks with solar or heliospheric phenomena. 

\begin{figure}
	\begin{center}
		\includegraphics[scale=0.5]{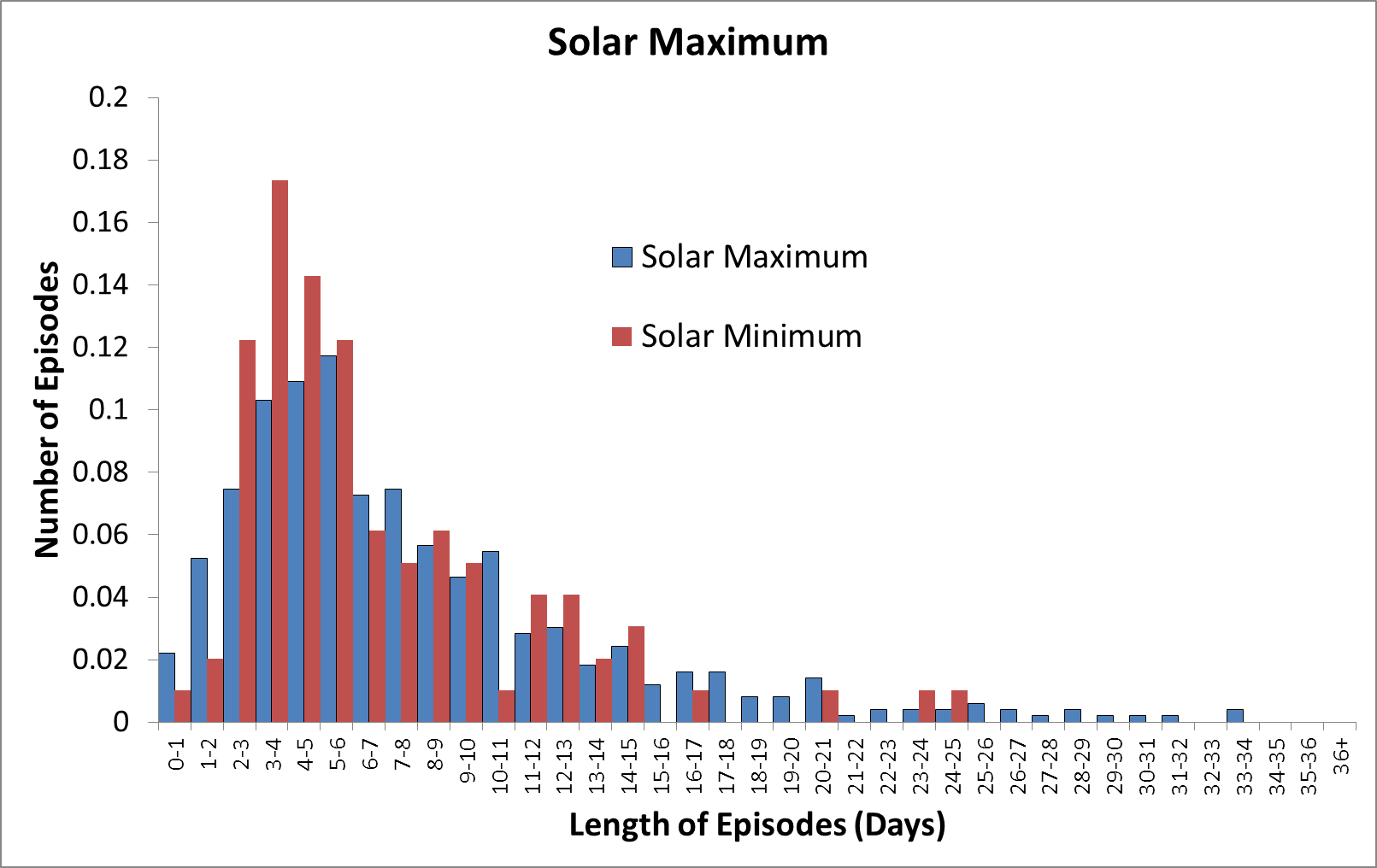}
		\caption{Distribution of episode lengths during solar maximum and solar minimum.}
		\label{maxmindist}
	\end{center}
\end{figure}

\begin{figure}
	\begin{center}
		\includegraphics[scale=0.45]{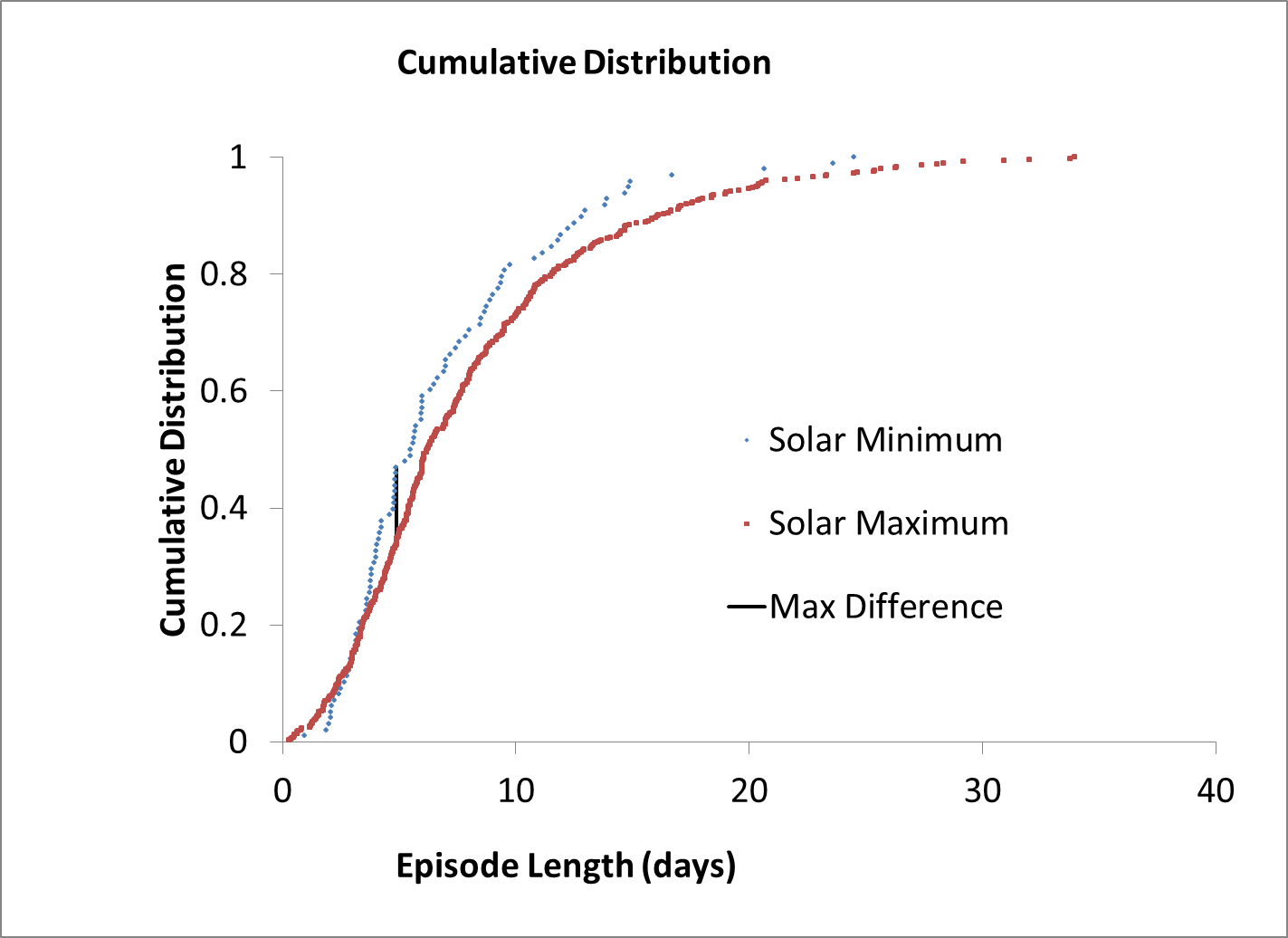}
		\caption{Cumulative distributions for episode length during solar maximum and solar minimum. The black vertical line is located at the point where the two distrubitions are the most different.}
		\label{maxminKS}
	\end{center}
\end{figure}

Episodes can last from less than one day to more than a month. The episodes were divided into those that occurred around solar maximum (between 2.5 years before the sunspot maximum and 4.5 years after the maximum) and the remaining episodes that occurred during solar minimum. The distributions of solar minimum and maximum episode lengths are shown in \figureref{maxmindist}. 
It can be seen that most frequent episode length during solar minimum was 3-4 days while during solar maximum it was 5-6 days, which suggests that these distributions might be different. To check this, these two distributions were compared using a two-sample Kolmogorov-Smirnov Test. The cumulative distributions used in this test are shown in \figureref{maxminKS}. 
The hypothesis that both distributions came from the same parent distribution could be rejected at a 90\% confidence level. This means that episode lengths tend to be longer during solar maximum.  This difference is expected because episodes are the result of overlapping solar proton events (SPE). Since SPEs are more frequent during solar maximum, overlaps should occur more often resulting in longer episodes, as was found.

\begin{figure}
	\begin{center}
		\includegraphics[scale=0.8]{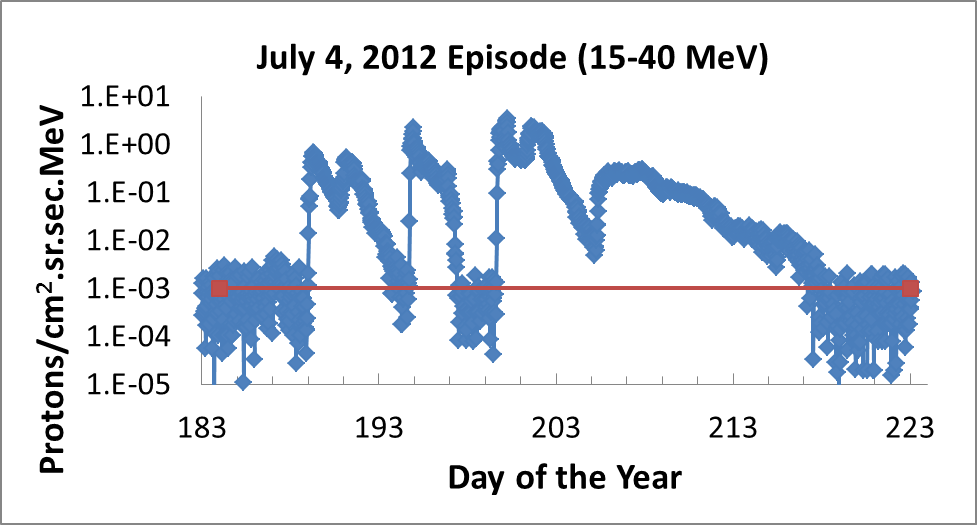}
		\caption{Channel P4 of the July 4, 2012 episode. The blue data points are the flux while the red line shows the background level in the channel.}
		\label{7412p4}
	\end{center}
\end{figure}

While longer episodes tend to have more peaks, a wide variation was found in the number of flux peaks during episodes with similar durations. Occasionally, episodes with multiple peaks will contain extended periods during which the flux remains nearly constant. Two examples of this can be seen in channel P4 during the episode that extends from July 4 to August 7, 2012, shown in \figureref{7412p4}. 
The 30-minute averaged flux varies by less than a factor of 2 for almost a day between days 201.4 and 202.3 and again for three days, between days 205.5 and 208.6. Such flux plateaus, lasting for a day or more, can appear in any energy channel.

Another feature of the episodes in this data set is the steepness of the rise and fall of the flux. Examples of this can also be seen in \figureref{7412p4}. The 30-minute-averaged flux increased in a few hours beginning in channel P4 on day 188.95 and again on days 194.43 and 199.41. There is also a rapid 5-hour decline in the flux in channel P4 between days 197.2 and 197.4. The rapid event onsets of these events indicate a direct magnetic connection between the Earth and the SEP acceleration site.

\begin{figure}
	\begin{center}
		\includegraphics[scale=0.8]{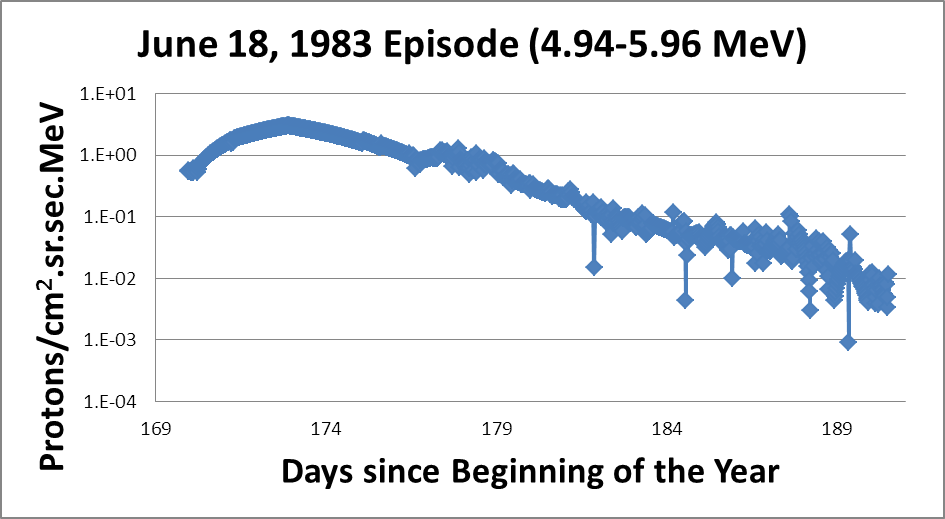}
		\caption{Channel 8 of the June 18, 1983 episode.}
		\label{61883c8}
	\end{center}
\end{figure}

Not all episodes contain sudden increases or decreases in the flux. In some episodes, the flux increases and decreases slowly. For example, the episode beginning on June 18, 1983 and lasting until July 9 is shown in \figureref{61883c8}. 
In this episode, the flux during the first event increases by more than a factor of 10 over a period of 2.8 days. During the second event in this episode (which occurs during the decline of the first), the flux increases more than a factor of two in 17 hours. Following the peak of the second event, the flux decreases slowly over a period of many days. The flux decreases more rapidly at higher energies. These events are typical of instances when the Earth is poorly connected to the acceleration site.  The 1983 and 2012 episodes cited here are the extremes. The other episodes exhibit a mix of rapidly and slowly varying fluxes corresponding to different degrees of connectivity between the Earth and the acceleration site. When an episode contains an event that is above background in channel P7, the event has a sudden onset in this channel. The higher energy particles of an SPE tend to arrive at the Earth quicker and are less affected by things that will slow down the arrival of the lower energy particles. 

\begin{figure}
	\begin{center}
		\includegraphics[scale=0.8]{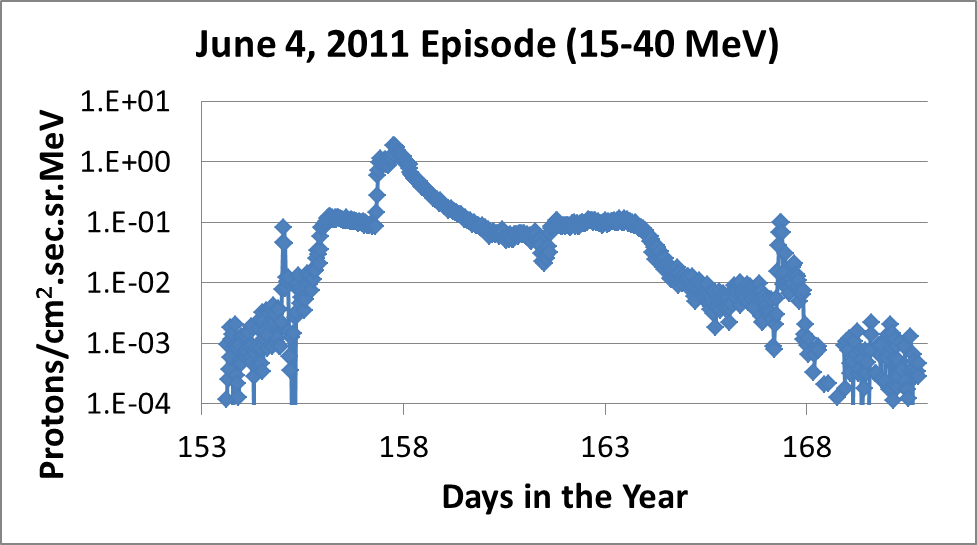}
		\caption{Channel P4 of the June 4, 2011 episode.}
		\label{6411p4}
	\end{center}
\end{figure}

The episode beginning on June 4, 2011 is shown in \figureref{6411p4}. 
This episode contains examples of events in which the flux declines steeply after it peaks. These are in striking contrast to the June 18, 1983 episode (\figureref{61883c8}), which has a decrease in the flux extending over a period of 15 days.

\begin{figure}
	\begin{center}
		\includegraphics[scale=0.8]{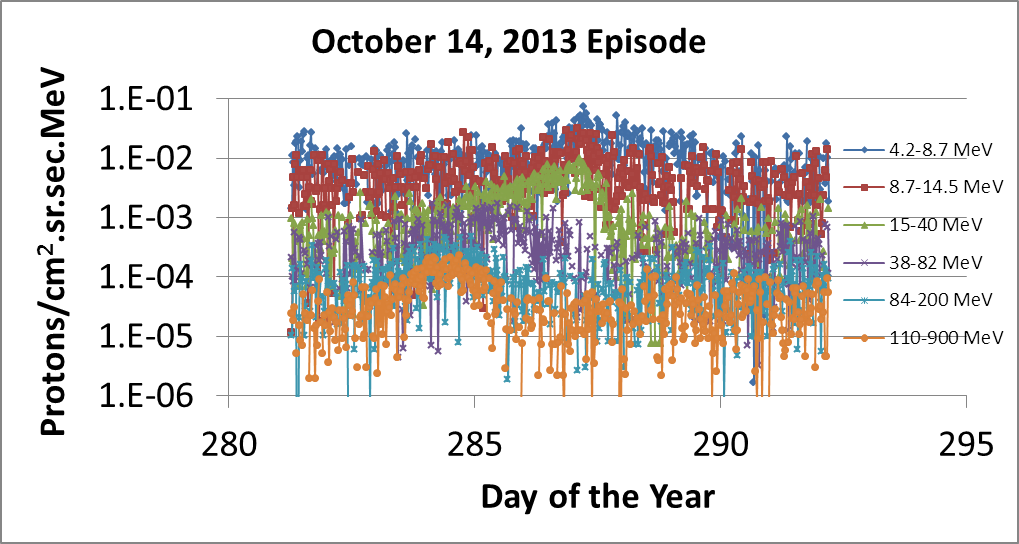}
		\caption{The October 14, 2013 episode.}
		\label{101413}
	\end{center}
\end{figure}

\begin{figure}
	\begin{center}
		\includegraphics[scale=0.8]{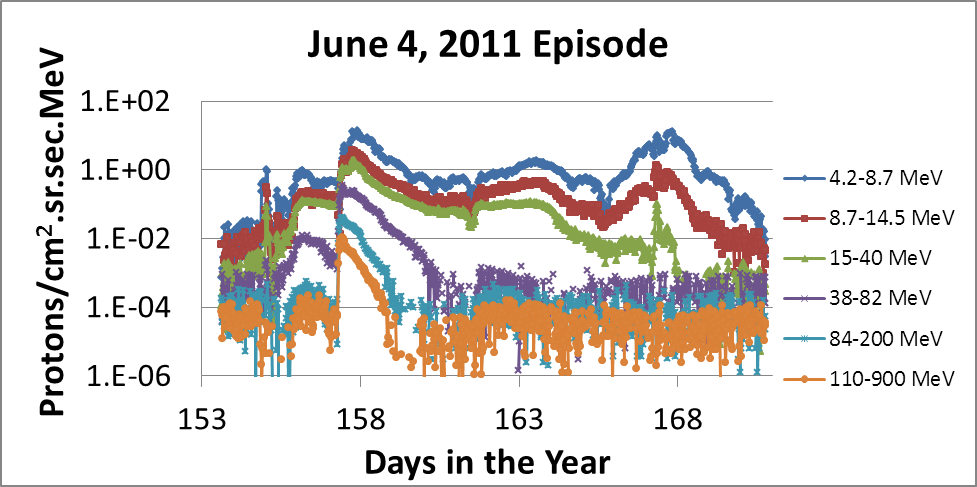}
		\caption{The June 4, 2011 episode.}
		\label{6411}
	\end{center}
\end{figure}

The variability in the hardness of the proton energy spectra is also apparent in this data set. A soft spectrum is one in which the flux declines rapidly with increasing energy while a hard spectrum declines slowly. An event with a soft spectrum is usually above background in the lowest energy channels only. A hard and soft spectra can be seen in the October 14, 2013 episode. The flux in each channel is shown in \figureref{101413}. 
The hard event can not be seen until channel P4. An episode with a hard spectrum usually contains a mixture of hard and soft events. There were a few episodes where the soft spectra contributed a significant portion of the total episode fluence in the lower energy channels and the hard spectra dominated the higher energy channels. In the June 4, 2011 episode (\figureref{6411}), the last peak is from a soft event that contributes slightly more than the first peak to the total episode fluence in channels P2 through P4. 
However, this spectrum is submerged in the background in channels P5 and above, whereas the first peak has a hard spectrum and is still well above background in channel P7.

This chapter has detailed the steps need to take the raw data from the satellites and process it into cleaned, background-subtracted data. This processed data was then used to identify episodes of increased proton flux measured around Earth. The flux was then converted to fluence and the GOES fluences were normalized to GME. Finally, the normalized GOES fluences were redistributed into GME energy channels to create a seamless database of episode fluences from 1973-2013 that spans the energy range 0.88-485 $MeV$. The chapter concluded with a description of some characteristics of episodes found in this database.

\chapter{Method and Results}
\label{method}
This chapter will discuss the extreme value theory that is used in the probabilistic model. The steps taken to find the cumulative distributions of fluences and the number of episodes per year used in extreme value theory are also described. The episode database developed in the previous chapter is used to create the cumulative distributions and to develop a model for the number of episodes per year. This chapter concludes with a section discussing the results of some preliminary testing of the probabilistic model.

\section{Extreme Value Theory}

The probabilistic model will create an upper bound proton spectrum to be used as a design reference environment. It can be tailored to a specific mission occurring during the time period 1953-2052. The model will use the mission start date, mission duration, and a user-specified confidence level to build the design reference environment at a distance of 1 $AU$ from the sun. The probabilistic model will use extreme value theory to build the reference environment for a specific mission. Extreme value theory gives the model a way to find the probability that the fluence from an episode will not exceed some limit during a mission. Following the work in \citet{Xapsos98}, the extreme value theory starts by maximizing the entropy, 
\begin{equation}
S=-\int_{0}^{M_{max}} p(M) \ln(p(M)) dM,
\end{equation}
where $S$ is the entropy, $p(M)$ is the probability density, and $M$ is defined as 
\begin{equation} \label{phi}
M=\log(\phi),
\end{equation}
where $\phi$ is the episode integrated fluence. If we subject the entropy equation to the conditions 
\begin{equation*}
	\int_{0}^{M_{max}} p(M) dM=1
\end{equation*}
and
\begin{equation*}
	\int_{0}^{M_{max}} M p(M) dM = \omega,
\end{equation*}
where $\omega$ is the mean of $M$ over the range in the integral, we can use the method of Lagrange multipliers to solve for $p(M)$ (This method also assumes that $M_{min}=0$ and there is a finite upper limit, $M_{max}$, whose value is unknown.). The calculations yield
\begin{equation}
p(M)=\frac{\lambda}{1-\exp(-\lambda M_{max})} \exp(-\lambda M),
\end{equation}
where $\lambda$ is a constant and a Lagrange multiplier. By integrating the probability density from $0$ to $M$, the cumulative probability can be found as
\begin{equation} \label{PM}
P(M)= \frac{1-\exp(-\lambda M)}{1-\exp(-\lambda M_{max})}.
\end{equation}
The initial distribution, $P(M)$, is the probability that the next solar particle episode will have a fluence $<\phi$. This equation can be applied to $n$ episodes in some period of time, $T$ years, which produces a probability, $[P(M)]^{n}$, that none of the $n$ episodes will have a fluence $\geq\phi$. If the average number of episodes per year with energy, $E$, is $\mu$, then Poisson's equation can be used to find the probability that $n$ episodes will be produced in $T$ years is
\begin{equation}
\frac{(\mu T)^{n} e^{-\mu T}}{n!}.
\end{equation}
The use of Poisson’s equation assumes that the episodes occur independently. This is essentially true for episodes, as was discussed in the beginning of \chapterref{data}. Now, the probability that no episode with a fluence $\geq\phi$ will occur in $T$ years is 
\begin{equation}
F_{T}(M) = \sum\nolimits_n \frac{(\mu T)^n}{n!}\exp(-\mu T) [P(M)]^n.
\end{equation}
The term $F_{T}(M)$ can also be thought of as the confidence level. It must be specified by the user of the probabilistic model. This equation for the confidence level can be simplified to
\begin{equation} \label{FT}
F_{T}(M) = \exp(-\mu T[1-P(M)]).
\end{equation}
Equations \eqref{phi}, \eqref{PM}, and \eqref{FT} can be solved for $\phi$ once the user specifies $T$ and $F_{T}(M)$.  This process can then be repeated for the different energy channels to create a differential energy spectrum for the episode-integrated fluence, of any element. A spectrum more intense than the one that was constructed in this way will not occur in $T$ years of the mission at a confidence level $F_{T}(M)$.

The probabilistic model will use the extreme value theory to construct the upper bound spectra for protons with the energy channels found in the episode database. Before the model can construct these bounding-case spectra, the cumulative distribution for each channel and the number of episodes per year are needed to solve Equations \eqref{phi}, \eqref{PM}, and \eqref{FT}.

\section{Cumulative Distributions}
\label{CumDist}
The cumulative distributions, $P(M)$, for each energy channel can be constructed using the episode database. All the cumulative distributions are really plotted as $1-P(M)$, meaning that the lowest fluence in a particular channel has a distribution value of 1 while the higher fluences are closer to 0. The cumulative distributions were plotted this way since the probabilistic model used the cumulative distributions in the same format. In each channel, there is a high frequency of episodes in the low fluence regime. This provides for good statistics in this part of the distribution, meaning that a line just has to be drawn through it to fit the data well. A power law and a log polynomial were chosen to fit this portion of the cumulative distributions. There are more statistical fluctuations in the high fluence regime, which means that the data has to be fit with a theoretical model. \citet{Xapsos98} showed that the higher fluence regime of the cumulative distributions can be fit with a Fr\'{e}chet distribution. These three equations were used to create a smooth cumulative distribution for each channel.

There are a few important points to consider when fitting these cumulative distributions. Each channel uses a different number of episodes to create the cumulative distribution for several different reasons. Softer and smaller episodes are not above background in the higher energy channels. This means that the cumulative distribution for each channel is constructed from a different number of episodes. The cumulative distribution for Channel 29 has roughly one sixth as many episodes in it as the lower fluence channels. Also, an episode was occasionally dropped from a single channel because its fluence in that channel was much lower than the other episodes. These outliers were removed because their low fluence appeared to be an artifact of the data analysis. This allowed for the cumulative distributions to be fit more accurately in the lower fluence range. 

\begin{figure}
	\begin{center}
		\includegraphics[scale=0.8]{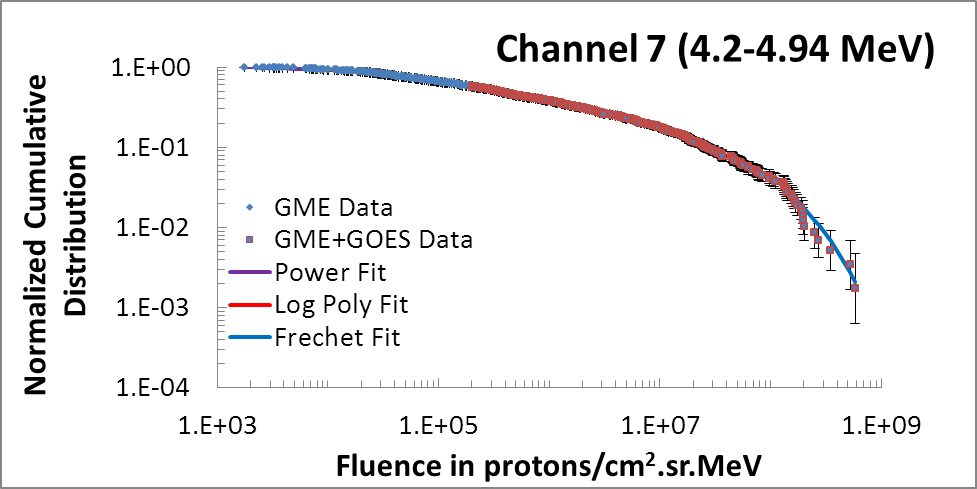}
		\caption{The cumulative distribution for channel 7.}
		\label{CumDistch7}
	\end{center}
\end{figure}

There were also episodes left out of the cumulative distributions for channels 6-11 (which correspond to the energy range associated with GOES channel P2). In \sectionref{fittingSpectra}, it was mentioned that episodes which were visible above background only in GOES channel P2 had to be thrown out since there was no way to accurately fit their spectra. If it is assumed that the GME and GOES episodes come from the same parent distribution, then the cumulative distribution for the GOES and GME episodes separately should be the same. Since some episodes were thrown out in these channels, the combined GME and GOES cumulative distribution will be undersampled in the lower fluence regime. This can be corrected for by using the GME only episodes until the episodes get above a fluence, which guarantees that the episode will be visible in more than one GOES channel. This was done by looking at the episodes recorded by GOES to find fluences in each of GME channels, 6-11, that were high enough to insure that episodes achieving this fluence level could not have been seen only in GOES channel P2. These fluences were then used as the threshold fluences for switching from the GME-only set of episodes to the combined set of GME and GOES episodes. The cumulative distributions for the combined GME and GOES episodes were then scaled to match the portions of the cumulative distributions created using the GME-only set of episodes. This does not distort the cumulative distributions and allows for more episodes to be used to better define the cumulative distributions in the higher fluence regime. \figureref{CumDistch7} shows the cumulative distribution for channel 7. 
The scaling factor used to scaled the GME and GOES data was 0.981 with a threshold fluence for switching of 2.0E5 $protons/cm^2.sr.MeV$.

\begin{figure}
	\begin{center}
		\includegraphics[scale=0.8]{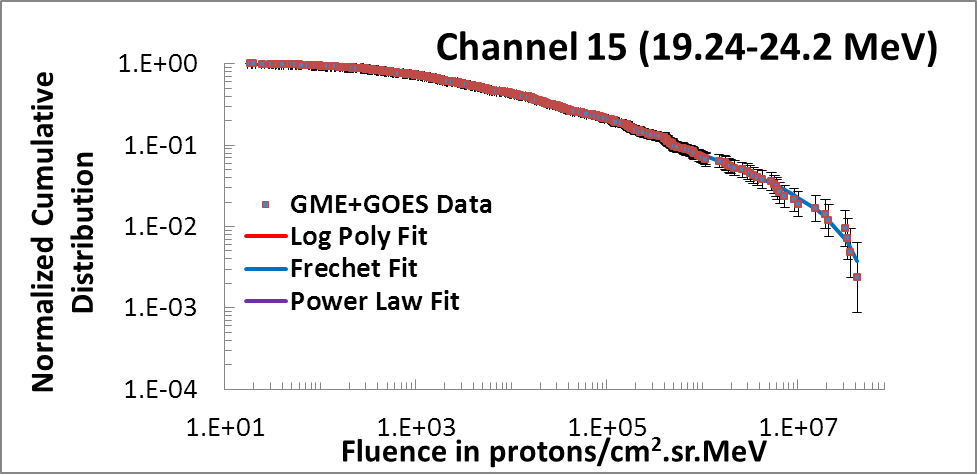}
		\caption{The cumulative distribution for channel 15.}
		\label{CumDistch15}
	\end{center}
\end{figure}

Now that the episodes for each channel have been identified, the next step is to fit the cumulative distributions with the equations. The first thing to do is to create the inverse logarithmic cumulative distribution, normalized to 1 at the lowest fluence. The equation for this inverse cumulative distribution is 
\begin{equation}
C_{I}=\frac{1}{1+\exp{[-\ln{\frac{n}{N-n}}]}},
\end{equation}
where $C_{I}$ is the inverse cumulative distribution, $N$ is total number of episodes used to determine the cumulative distribution. If episode are sorted from largest to smallest, then the sequence number of the smallest is also $N$. The sequence number of an episode is $n$. The sequence number is less than or equal to $N$ and greater than or equal to 1. Before fitting $C_{I}$ can begin, error bars need to be created for each episode in $C_{I}$. The error bars for a cumulative distribution were found following the work of \citet{MacKay06}. The standard deviation of the inverse cumulative distribution is
\begin{equation}
\sigma_{SD}=\sqrt{\frac{N}{n(N-n)}},
\end{equation}
where $\sigma_{SD}$ is the standard deviation. Using $\sigma_{SD}$ and $C_{I}$, the upper and lower error bars can be found for each episode in the channel. The equation,
\begin{equation}
\sigma_{\pm}=\left|\frac{1}{1+\exp{[-\ln{\frac{n}{N-n}}]\mp \sigma_{SD}}} - C_{I} \right|,
\end{equation}
can be used to find the upper and lower error bars, respectively. Now, the inverse cumulative distributions can be fitted with the three equations mentioned earlier. For each channel, the Fr\'{e}chet distribution was fitted to the upper fluences and extended to the lower fluences as much as possible. As can be seen in \figureref{CumDistch15}, the Fr\'{e}chet distribution was extended to the lower fluences until it compromised the fit of the higher fluences, using the error bars as a guide. 
The power law was used to fit the lowest fluences since the data are basically a straight line at the start of the cumulative distribution when plotted on a log-log graph. The middle section of the cumulative distribution is fitted with a log polynomial of sixth order. This polynomial was fitted using the trendline function in EXCEL. The log polynomial fit of the remaining data, combined with the power law and Fr\'{e}chet fits, gives a fit for the entire cumulative distribution that fell within the error bars for each data point. These three functions were then adjusted to give the best fit to the cumulative distribution. The adjustments made to them were moving the starting and ending fluence for each equation in the fit and adjusting the equations to make sure that the cumulative distribution was a smooth function. These adjustments were done together to find the best fit to the cumulative distribution. The cumulative distribution fits for all 29 channels can be found online \citep[see][]{cumdistfitsfile}.

\section{Episodes Per Year}
\label{episodes}
The number of episodes per year used for the model comes from three different sources: a sunspot proxy, actual number of episodes, and an eleven year cycle fit. Using the 40-year episode database and the catalog of the smoothed monthly sunspot numbers, a prediction tool was developed to calculate the number of episodes per year from the sunspot numbers. This tool, or proxy, allows the probabilistic model to calculate the number of episodes per year in 1954 to 1973 and 2014 to 2019. The actual number of episodes per year discovered in the dataset was used for the years 1974 to 2013. Finally, the 11-year cycle fit was used to calculate the episodes per year for 2020 to 2053.

Sunspots have been observed and counted going back into the 1600's. At the present moment, sunspots are the only characteristic of the sun that can be predicted with a high level of accuracy. \citet{Hathaway94} have been able to predict the monthly sunspot numbers accurately. The only concern regarding this prediction is that it is limited to predicting the future of the current solar cycle. Once the solar cycle starts, this tool can predict the monthly sunspot numbers for the entire cycle. But it can not predict further into the future. When the solar cycle is approaching its end, this tool can not predict any further into the future than it could at the beginning of the current cycle. Currently, this tool can predict the monthly sunspot numbers until the end of 2020.

\begin{figure}
	\begin{center}
		\includegraphics[scale=0.7]{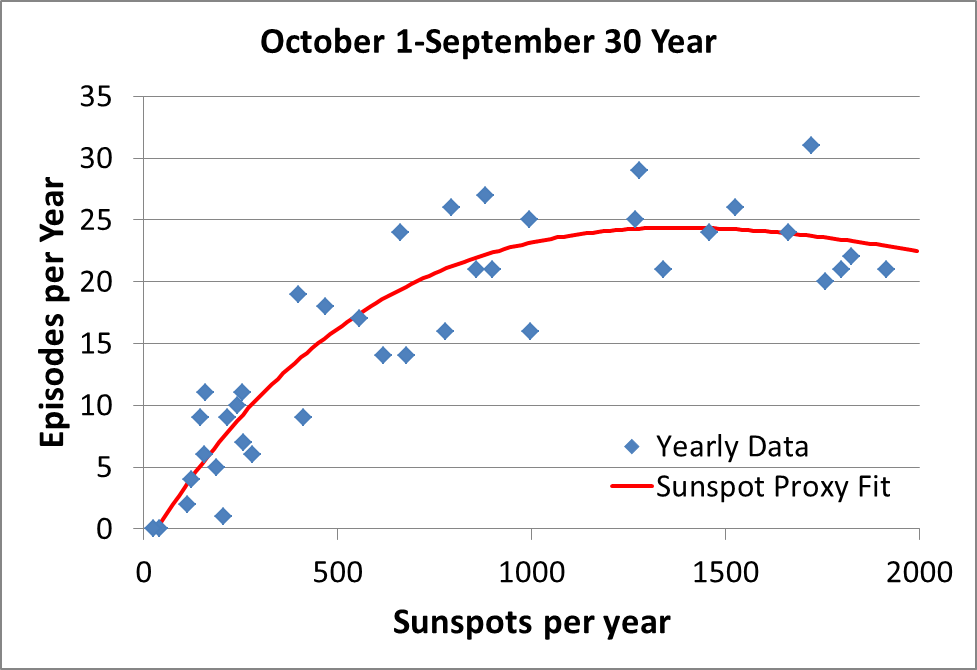}
		\caption{The sunspot proxy fit. Notice that as there are more sunspots in a year, the number of episodes per year tends to get smaller.}
		\label{SunspotFit}
	\end{center}
\end{figure}

\begin{figure}
	\begin{center}
		\includegraphics[scale=0.6]{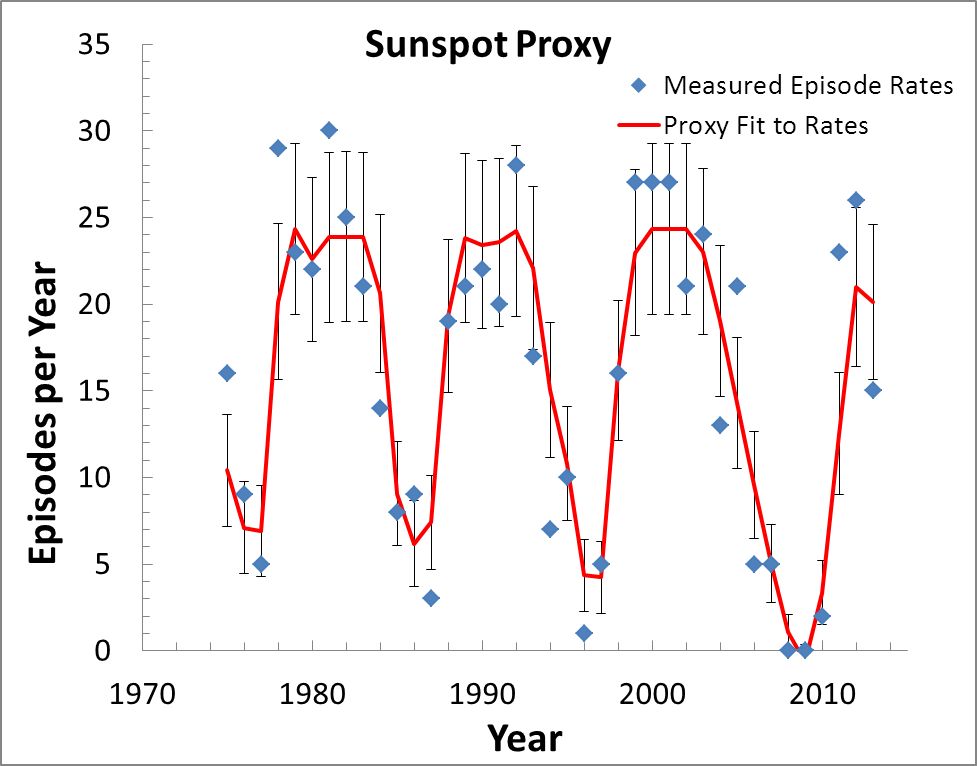}
		\caption{The number of observed episodes per year is compared to the number of episodes per year found using the sunspot proxy.}
		\label{SunspotComp}
	\end{center}
\end{figure}

To create a sunspot proxy to determine the number of episodes per year, a database of the smoothed monthly sunspot number needed to be obtained. The smoothed monthly sunspot numbers were downloaded off the NOAA website for 1974 to the present \citep[see][]{sunspot}. To change the monthly sunspot number into yearly sunspot numbers, twelve consecutive months were added together. Twelve sets of yearly data were compiled with each set having the year start on a different month, i.e. January-December, February-January, etc. Since there was not a full year of data for 1973 and to minimize human bias, the best fitting dataset out of the twelve would be used for the sunspot proxy. To find a sunspot proxy, the number of episodes per year and the number of sunspots per year were plotted for one of the twelve datasets. A few different fits were tried, including linear and quadratic functions. The one producing the best chi squared value was the exponential distribution with a dead time correction factor,
\begin{equation}
N = (a \ast n + b) \exp[-q(a \ast n +b)]
\end{equation}
where $N$ is the number of episodes in a year, $n$ is the number of sunspots in a year, and $a$, $b$ and $q$ are fitted parameters. This equation was then fitted by the other eleven datasets and the chi squared values for each fit were found. These values can be found in \tableref{Chisunspot}. 
\begin{table}
	\begin{center}
		\caption{The Chi squared values for the sunspot proxy.}
		\label{Chisunspot}
		\begin{tabular}{|c|c|c}
			\cline{1-2}
			\multicolumn{2}{|c|}{Year} & \\ \hline
			Start & End & \multicolumn{1}{|c|}{Chi Squared} \\ \hline
			January & December & \multicolumn{1}{|c|}{0.84} \\ \hline
			February & January & \multicolumn{1}{|c|}{0.92} \\ \hline
			March & February & \multicolumn{1}{|c|}{1.07} \\ \hline
			April & March & \multicolumn{1}{|c|}{1.20} \\ \hline
			May & April & \multicolumn{1}{|c|}{1.19} \\ \hline
			June & May & \multicolumn{1}{|c|}{1.18} \\ \hline
			July &  June & \multicolumn{1}{|c|}{1.25} \\ \hline
			August & July & \multicolumn{1}{|c|}{1.03} \\ \hline
			September & August & \multicolumn{1}{|c|}{1.07} \\ \hline
			October & September & \multicolumn{1}{|c|}{1.31} \\ \hline
			November & October & \multicolumn{1}{|c|}{0.97} \\ \hline
			December & November & \multicolumn{1}{|c|}{0.84} \\ \hline
		\end{tabular}
	\end{center}
\end{table}
The chi squared values were all basically around 1, which corresponds to a perfect fit. The fitted values for the year starting October 1 and ending September 31 of the following year were chosen to be used for the probabilistic model due to the agreement with the 11-year cycle, which will be discussed later in this section. The fitted parameters of the sunspot proxy are shown in \tableref{proxy}.
\begin{table}
	\begin{center}
		\caption{The fitted parameters of the sunspot proxy.}
		\label{proxy}
		\begin{tabular}{|c|c|}
			\hline
			Parameter & Value \\ \hline
			$a$ & 0.0491 \\ \hline
			$b$ & -1.6122 \\ \hline
			$q$ & 0.0151 \\ \hline
		\end{tabular}
	\end{center}
\end{table}
The sunspot proxy fit can be seen in \figureref{SunspotFit}. 
\figureref{SunspotComp} shows the sunspot proxy compared to the actual observed number of episodes in each year in this episode database. 
This equation also makes sense when used for the number of episodes per year. An active sun produces more sunspots than a quiet sun. Since SPEs are more frequent when the Sun is active, the SPE frequency can be expected to correlate with the sunspot number. Initially, as the number of events per year is increased, the number of episodes per year increases linearly. However, there comes a point where an increase in the number of events per year causes the number of episodes per year to grow less rapidly before it eventually even decreases. This is caused by the fact that when more events occur during a year, the events start to overlap more frequently. The increase in the number of events merges events into longer episodes. Between 1954-1973, the sunspot proxy was used to estimate the number of episodes in a given year since there is data on the actual number of sunspots recorded during this time. This gives a value for the number of episodes per year based on the only consistently measured property of the sun. 

For the years 1973-2013, the actual number of episodes per year was used in the probabilistic model. To determine the actual number of episodes in a given year, the episodes were tallied up by looking at the starting date of the episode. Whichever month the episode started in, the episode was said to occur in that month no matter even if the episode started on the last day of the month. This is not a perfect method, i.e. the October 26, 2013 episode, but this eliminated any human bias in determining which month the episode should be attributed to when the episode spanned two months. 

The sunspot proxy was used for the years 2014-2019. As mentioned earlier, sunspots can be predicted into the future with high accuracy once the cycle nears maximum. The current solar cycle, Solar Cycle 24, began in 2008 which allows for reliable predictions of sunspots through the end of the cycle (roughly 2020). These predictions can be found on NOAA's website \citep[see][]{sunspotpredict}. For the probabilistic model to be as accurate as possible, the model needs to use the most accurate predictions available. Since there is no record of how often these predicts may change, it is recommended to recalculate the episode predictions using the sunspot proxy every few months.

\begin{figure}
	\begin{center}
		\includegraphics[scale=0.8]{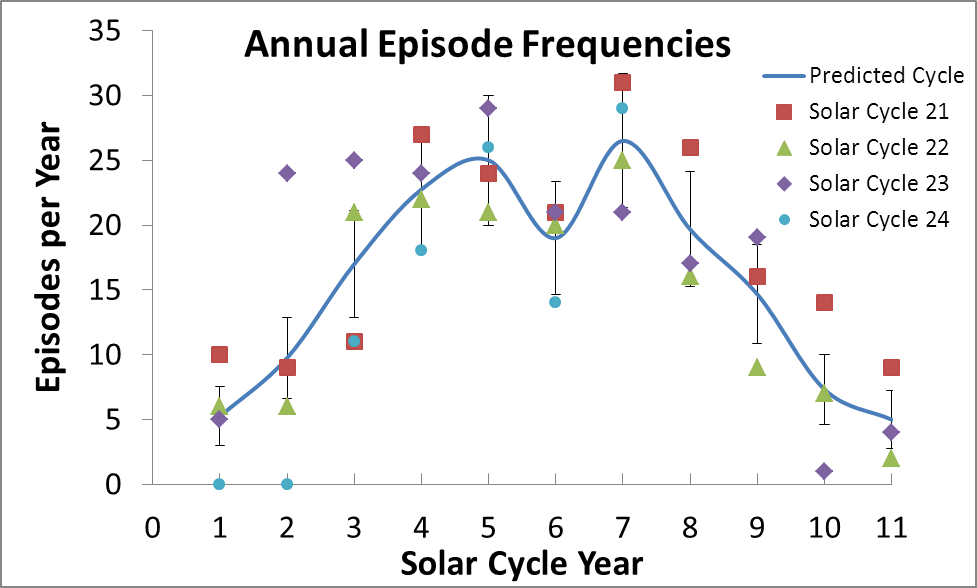}
		\caption{The 11-year cycle fit is compared to the actual number of episodes per year in each solar cycle.}
		\label{11cycle}
	\end{center}
\end{figure}

For the years 2020-2053, an 11-year solar cycle fit was used to determine the number of episodes per year. An average solar cycle lasts about 11 years so this length was used for this fit. Since the solar cycles fluctuate in length from cycle to cycle, the probabilistic model was extended only to 2053. It was decided that the 11-year cycle fit was not to be used to extend the model even further because of the chance that the model would fall out of synchronization with the solar cycle and would become much less accurate. To create the 11-year solar cycle fit to determine the number of episodes per year, the number of episodes per month were counted up. Twelve datasets were once again created with each dataset starting at a different month of the year. The RMS value and the reduced chi squared were found for all twelve sets of data. The set with the lowest Reduced Chi Squared was the data set used for the 11-year cycle. This turned out to be the year starting on October 1 and ending on September 30. This dataset was substantially better than the other 11 sets, as shown in \tableref{Chi11cycle}.  
\begin{table}
	\begin{center}
		\caption{The Chi squared values for the 11-year solar cycle fit.}
		\label{Chi11cycle}
		\begin{tabular}{|c|c|c}
			\cline{1-2}
			\multicolumn{2}{|c|}{Year} & \\ \hline
			Start & End & \multicolumn{1}{|c|}{Chi Squared} \\ \hline
			January & December & \multicolumn{1}{|c|}{3.11} \\ \hline
			February & January & \multicolumn{1}{|c|}{3.13} \\ \hline
			March & February & \multicolumn{1}{|c|}{3.20} \\ \hline
			April & March & \multicolumn{1}{|c|}{3.36} \\ \hline
			May & April & \multicolumn{1}{|c|}{3.24} \\ \hline
			June & May & \multicolumn{1}{|c|}{3.03} \\ \hline
			July &  June & \multicolumn{1}{|c|}{3.35} \\ \hline
			August & July & \multicolumn{1}{|c|}{3.39} \\ \hline
			September & August & \multicolumn{1}{|c|}{3.09} \\ \hline
			October & September & \multicolumn{1}{|c|}{2.81} \\ \hline
			November & October & \multicolumn{1}{|c|}{3.03} \\ \hline
			December & November & \multicolumn{1}{|c|}{3.12} \\ \hline
		\end{tabular}
	\end{center}
\end{table}
The 11-year solar cycle fit is compared to the previous solar cycles in \figureref{11cycle}. 
There were some months were left over because they would not fit into a whole year. These extra months were added into the 11-year solar cycle fit for the year starting on October 1. This slightly adjusted the values of two years in the solar cycle fit. The last step was to adjust the start of the cycle so that the years with the fewest episodes were at the beginning and end of the cycle. Since the sunspot proxy continues until the end of the current solar cycle, the 11-year cycle needs to start at solar minimum. Since this was not possible with the fit due to an increase in the number of episodes per year in the last year of the fit, the solar cycle was chosen to start at the point where there are the fewest episodes per year. When the 11-year cycle is adjusted, this ensures that the cycle and sunspot proxy are both in phase with the solar cycle when the probabilistic model switches from the sunspot proxy to the 11-year cycle to calculate the number of episodes per year. The solar cycle fit can be seen in \tableref{solarcyclefit}.
\begin{table}
	\begin{center}
		\caption{The 11-year solar cycle fit}
		\label{solarcyclefit}
		\begin{tabular}{|c|c|}
			\hline
			Cycle Year & Episodes \\ \hline
			1 & 5.25 \\ \hline
			2 & 9.75 \\ \hline
			3 & 17.00 \\ \hline
			4 & 22.75 \\ \hline
			5 & 25.00 \\ \hline
			6 & 19.00 \\ \hline
			7 & 26.50 \\ \hline
			8 & 20.60 \\ \hline
			9 & 14.67 \\ \hline
			10 & 7.33 \\ \hline
			11 & 8.17 \\ \hline
		\end{tabular}
	\end{center}
\end{table}

A program has been created that generates a list containing the number of episodes per year for each year in the probabilistic model. This program updates the file containing the yearly episode numbers, which is required to run the probabilistic model, while making sure that the file is in the correct format for input to the model. This program will also update the number of episodes per year based on the sunspot predictions if the predictions have changed on the NOAA website. This program will work until the end of the solar cycle or until the format of the website where the sunspot predictions are stored changes. This program has been included in \appendixref{episodeCalc}.

\section{Testing the Model}

\begin{figure}
	\begin{center}
		\includegraphics[scale=.9]{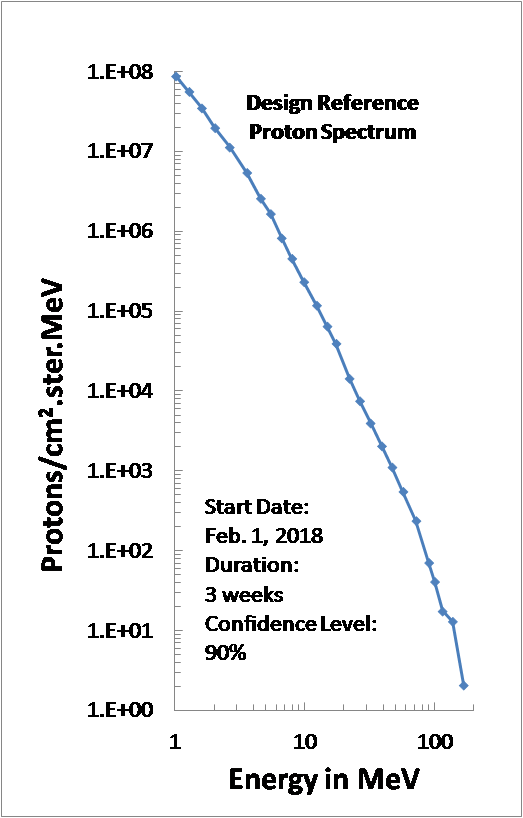}
		\caption{The upper bound proton reference environment for a series of hypothetical EVAs for astronauts on the ISS that begins on February 1, 2018.}
		\label{DRPE2118}
	\end{center}
\end{figure}

\begin{figure}
	\begin{center}
		\includegraphics[scale=1]{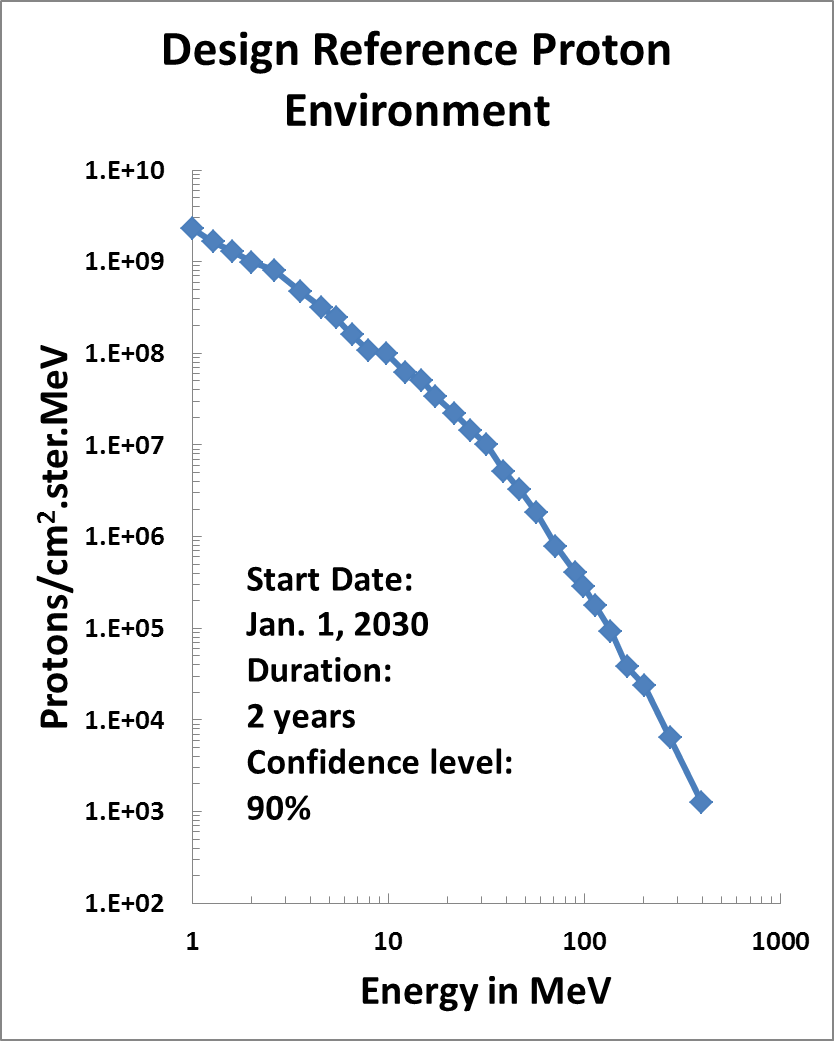}
		\caption{The upper bound proton reference environment for a hypothetical satellite mission that begins on January 1, 2030.}
		\label{DRPE1130}
	\end{center}
\end{figure}

With all the pieces of the probabilistic model finished, the model can now be assembled and tested. The latest version of the probabilistic model can be found in \appendixref{probabilistic}. Two hypothetical missions were planned and graphed using this program. The first one is a short series of EVAs for the astronauts on the International Space Station (ISS). This mission will start on February 1, 2018 and will last three weeks. The confidence level is set for 90\%. This upper bounding spectrum for this mission can be seen in \figureref{DRPE2118}.
The second is a proposed mission for a satellite that will orbit the sun at 1 $AU$. This mission will begin on January 1, 2030 and will last 2 years. The mission is set with a 90\% confidence level. \figureref{DRPE1130} shows the upper bounding spectrum for this mission.

\begin{figure}
	\begin{center}
		\includegraphics[scale=0.6]{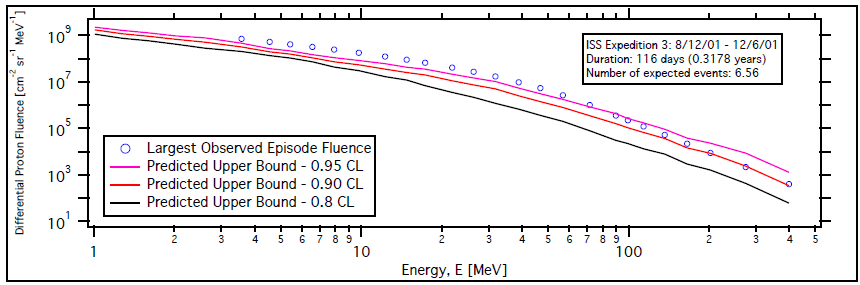}
		\caption{The largest episode fluence is graphed alongside the three predictions from the probabilistic model for ISS Expedition 3. This was the only expedition that contained an episode whose fluence exceeded the predictions of the probabilistic model at the 95\% confidence level.}
		\label{iss3}
	\end{center}
\end{figure}

The probabilistic model was tested by identifying 20 ISS expeditions from the time period 2000 to 2012 and comparing the largest episode fluence seen during each mission to the upper bounding spectrum produced by the probabilistic model. The 20 missions chosen are listed in \tableref{TestSpec}.
\begin{table}
	\begin{center}
		\caption{The 20 ISS expeditions start and end dates used to test the Probabilistic Model.}
		\label{TestSpec}
		\begin{tabular}{|c|c|c|}
			\hline
			ISS Expedition & Start Date & End Date \\ \hline
			1 & 10/31/00 & 3/9/01 \\ \hline
			2 & 3/10/01 & 8/11/01 \\ \hline
			3 & 8/12/01 & 12/6/01 \\ \hline
			4 & 12/7/01 & 5/13/02 \\ \hline
			5 & 6/5/02 & 12/7/02 \\ \hline
			7 & 5/26/03 & 10/28/03 \\ \hline
			8 & 2/11/04 & 4/20/04 \\ \hline
			9 & 4/21/04 & 10/15/04 \\ \hline
			10 & 10/16/04 & 4/16/05 \\ \hline
			11 & 4/17/05 & 10/2/05 \\ \hline
			13 & 4/1/06 & 9/17/06 \\ \hline
			14 & 9/18/06 & 5/14/07 \\ \hline
			23 & 4/4410 & 6/6/10 \\ \hline
			24 & 6/17/10 & 10/8/10 \\ \hline
			26 & 12/17/10 & 05/06/11 \\ \hline
			27 & 4/4/11 & 9/16/11 \\ \hline
			28 & 6/7/11 & 11/22/11 \\ \hline
			29 & 11/14/11 & 4/27/12 \\ \hline
			30 & 12/21/11 & 7/1/12 \\ \hline
			31 & 5/5/12 & 9/17/12 \\ \hline
		\end{tabular}
	\end{center}
\end{table}
For each expedition, the probabilistic model was run at a 80\%, 90\%, and 95\% confidence level. The resulting spectra were compared to the largest episode fleunce seen during the mission. The bounding spectrum created by the probabilistic model at the 95\% confidence level was higher than the worst episode fluence seen during the mission in 19 out of the 20 expeditions. This result is exactly what is expected from using the probabilistic model at a 95\% confidence level. A more in-depth look at these results will allow the model user to get a better feel for how well the upper bound spectrum will work for a specific mission. ISS Expedition 3, shown in \figureref{iss3}, is the only expedition that exceeded the probabilistic model at the 95\% confidence level.
It's interesting to note that the 95\% confidence level prediction was exceeded only in the lower energy regime. The model was successful for the higher energy regime. The spectra for the other two predictions were very close to the actual fluence observed during this episode. 

\begin{figure}
	\begin{center}
		\includegraphics[scale=0.6]{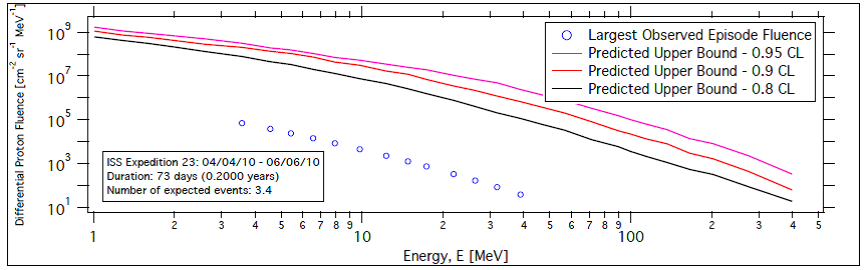}
		\caption{ISS Expedition 23 shows the worst overprediction of the probabilistic model.}
		\label{iss23}
	\end{center}
\end{figure}

\begin{figure}
	\begin{center}
		\includegraphics[scale=0.6]{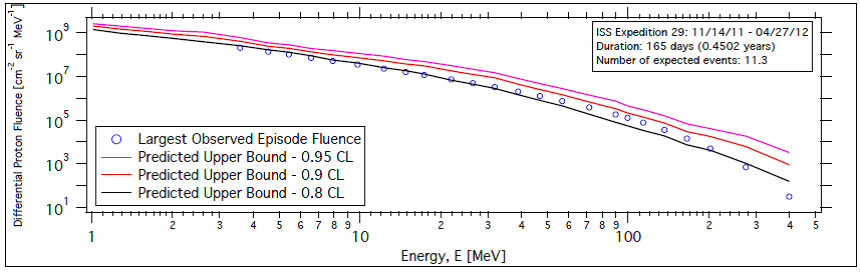}
		\caption{The probabilistic model was successful at the 90\% and 95\% confidence level for ISS Expedition 29.}
		\label{iss29}
	\end{center}
\end{figure}

There were a few expeditions where the probabilistic model greatly overpredicted the upper bounding spectrum. The worst overprediction was for ISS Expedition 23, shown in \figureref{iss23}.
There are a couple reasons why the predicted upper bounding spectrum was so high for this expedition. First, there was only one episode that occurred during this expedition instead of the 3.4 that were predicted. The more episodes that occur during a mission, the higher the predicted upper bounding spectrum will be for the mission. Since the probabilistic model uses the number of episodes during the mission to create the spectrum, an incorrect number of episodes will cause the spectrum to be less precise. Second, the episode that occurred during this mission was very small. Out of the 20 expeditions studied, the episode in this mission had the smallest observed fluence by close to a factor of 10. Finally, this mission occurred during a quiet period in solar activity. This means that there was less activity occurring on the sun during this mission. This can be a reason why the observed episode fluence for this mission is very small. It should be noted that most of the predicted upper bound spectra that were much higher than the observed episode fluences occurred during the quiet period towards the end of solar cycle 23 and the beginning of solar cycle 24. For ISS Expedition 29 (\figureref{iss29}), the probabilistic model was successful at the 90\% and 95\% confidence level but fell short at the 80\% confidence level.

In this chapter, the probabilistic model was built using extreme value theory. The cumulative distributions for each channel were constructed and a model was developed to create the number of episodes per year. The cumulative distributions and the number of episodes per year are both required to construct the bounding-case spectra from the extreme value theory used in the probabilistic model. The probabilistic model was then tested using the ISS expeditions to determine the model's accuracy compared to these historical missions.

\chapter{Conclusion}
\label{conclusion}
A new model for predicting the proton environment at Earth has been discussed in this thesis. The model uses extreme value theory to build a mission-specific, upper bound spectrum at 1 $AU$ from the sun for episodes of elevated proton radiation levels. The model will allow the user to choose a mission start date, mission duration, and a confidence level in order to build a mission specific spectrum for a mission occurring during the time period 1953-2052. The model also uses episode fluence rather than the more traditionally used event fluence. A new database of episode fluences was created for the probabilistic model that spans the energy range 0.88-485 $MeV$.

The probabilistic model discussed in the previous chapters will allow spacecraft designers and mission planners to have a predicted design reference spectrum of episode-integrated proton fluence for their particular mission. The model works for missions ending before the year 2053. With the predictive capabilities that far in advance, the model can be used in some of the largest and most time consuming missions. This model can be used by spacecraft designers to determine the exact amount of shielding that will be needed for a mission before the proposed budget is selected and fully funded. This will allow mission planners to budget and plan the correct amount of radiation shielding for their mission. Since each mission is unique in the length and duration, each mission requires a different amount of shielding. If a mission is going to last a month during solar minimum, the design reference environment will be much different than a mission lasting 4 years during solar maximum. The mission during solar minimum would require much less shielding than the solar maximum mission because of phase of the solar cycle. If the solar maximum mission was designed with the shielding of the solar minimum mission, there would not be enough shielding to protect mission instruments or astronauts from the higher levels of radiation the mission would encounter. But overdesigning can be just as bad for the mission. In the current budgetary environment, the ability to minimize the costs of a mission could make the difference between acquiring funding or being rejected. A mission would waste resources if the radiation shielding was designed to protect against the worst episode of solar activity ever observed when the mission duration is not very long. The chance of some of the most massive storms occurring during a 3 week mission is very unlikely.

The database of episode fluences discussed in this thesis is currently the largest database of this kind, if not the only one. Since most databases and models have always used events, this new database will allow for other models to start using episodes instead of events. As stated earlier in this thesis (\sectionref{dataprocessingsection}), events cannot always be distinguishable from one another. This can make it difficult to create a database of isolated events. Since this new database uses episodes, each entry in the database will be more independent than a database of events. This database also has one of the wider energy ranges and has the finest energy binning of any database. This allows for the database to be more precise. The spectra produced by the model in this thesis is comprised of 29 data points, which allows for more accuracy throughout the spectrum. 

There are further areas of research related to this thesis that could be undertaken. First, the database of episode fluences could be expanded in the future with each passing year. This will increase the size of this database and keep it relevant moving into the future. Second, the time evolution of episodes can be examined and modeled. This will give mission planners a tool to determine if an ongoing event or episode on the sun poses serious threats to satellites or astronauts. Being able to predict whether the proton flux around the Earth will increase fast or slow gives the mission planners better information to determine what steps need to be taken to provide the most protection for the satellites and space crews. Third, the model can be expanded to calculate the proton fluence at different distances from the sun, not just 1 AU. With mission being planned to visit other planets, it is important to be able to predict the proton fluences at different places in the solar system and not just at Earth. 

In this thesis, a new probabilistic model for protons for the time period 1953-2052 was discussed. This model uses episode fluences to build a proton spectrum at a distance of 1 $AU$ from the sun, which can be designed to a user specified mission and confidence level. The resulting design reference spectrum has 29 data points that span the energy range 0.88-485 $MeV$. The episode fluences come from a newly created database from IMP-8 and GOES satellites. The database contains episodes from November 1, 1973 to December 31, 2013, making it the largest seamless database for proton fluences that currently exists. 

\begin{appendices}
\appendix

\chapter{The Probabilistic Model}
\label{probabilistic}
This appendix contains the python code that was used to create the probabilistic model that is discussed in this thesis. The probabilistic model has two input files. The first is the file containing the number of episodes per year. \appendixref{episodeCalc} contains the program that will produce this file. The second input file is the .csv file that contains all the cumulative distribution fits that were described in \sectionref{CumDist}. This file can be found online \citep[see][]{cumdistfitsfile}. Please note that the minimum mission duration that can be used in the probabilistic model is two weeks.

\singlespacing
\begin{lstlisting}
# Program to calculate the Bounding-Case differential fluence spectrum
# of solar Energetic Protons
# The Bounding-Case Spectrum is calculated for:
#	1) a user-specified confidence level
#	2) The mission launch date and duration specified by the
# user
#
import csv
import math
#
# First define the directory addresses of the CCMC directory on
# various computers used to develop this program
#
goesaddlst=list()
goesaddlst.append('C:\\Documents and Settings\\Jim\\My Documents\\Dropbox\\Probabilistic SPE Model\\CCMC\\')
goesaddlst.append('C:\\Users\\Jim Adams\\Dropbox\\Probabilistic SPE Model\\CCMC\\')
goesaddlst.append('C:\\Documents and Settings\\zach\\Desktop\\Dropbox\\Probabilistic SPE Model\\CCMC\\')
goesaddlst.append('C:\\Users\\jhadams1\\Dropbox\\Probabilistic SPE Model\\CCMC\\')
#
#Name the files to be opened.
#
nfl='cumulative fits.csv'
mep='Mean_Annual_Episode_Frequency.csv'
#
#Choose the first trial address.
#
address=goesaddlst[0]
#
# Concatenate the the filename to the first trial address.
#
infil=address+nfl
#
#test the address to see if it works.
#
try:
dfn = open(infil, 'r')
except IOError:
#
#Trial address did not work, construct and test the next one
#
address=goesaddlst[1]
infil=address+nfl
try:
dfn = open(infil, 'r')
except IOError:
address=goesaddlst[2]
infil=address+nfl
try:
dfn = open(infil, 'r')
except IOError:
address=goesaddlst[3]
infil=address+nfl
try:
dfn = open(infil, 'r')
except IOError:
print "None of the directories contain",nfl
print "Edit the PYTHON program to add the path to the directory containing",nfl
exit(0)
dfn.close()
inmep=address+mep
#
#--------------------------------------------------------
#
# Begin to read in the cumulative distribution fit
# parameters and calculate the cumulative fits.
#
# Find the number of channels in 'cumulative fits.csv'.
#
nch=len(open(infil,'r').readlines())
print 'Number of Channels = ', nch
#
# Read 'infil' as a .csv file, so that each line is a list
#
read = csv.reader(open(infil))
#
# Create 'lines' as a list to contain the lists in the lines of
# infil
#
lines=list()
#
# Append the lines in 'infil' to the list.
#
for line in read:
lines.append(line)
#
# Define lists to contain the cumulative distribution fit
# values (pch) and the corresponding fluences (fch).
#
pch=list()
fch=list()
#
# Define a list to hold the scale factors for the mean
# number of episodes during the mission in each energy
# channel.
#
Scale=list()
#
# Each line of cumulative fits.csv contains the
# parameters for fitting the cumulative distribution
# in one channel. Extract the fit parameters from each
# line of the input file.
#
for j in range(nch):
# Columin A contains the mean bin energy
# a through d are the Frechet parameters
a=float(lines[j][1]) #Column B
b=float(lines[j][2]) #Column C
c=float(lines[j][3]) #Column D
d=float(lines[j][4]) #Column E
# Fluence is where the Frechet fit starts and
# the log Polynomial fit ends.
Fluence=float(lines[j][5]) #Column F
# fmin and fmax are the smallest and largest
# fluence in the energy channel.
fmin=float(lines[j][6]) #Column G
fmax=float(lines[j][7]) #Column H
# Coef is a list of the seven coefficients of
# the log polynomial fit.
coef=list()
for k in range(7): #Columns I through O
coef.append(float(lines[j][8+k]))
# pflu is where the power-law fit ends and
# the log polynominal fit begins.
pflu=float(lines[j][15]) #Column p
# Amp and Index are the power law parameters.
Amp=float(lines[j][16]) #Column Q
Index=float(lines[j][17]) #Column R
# Sc is the scaling factor for mean episode frequencies
Sc=float(lines[j][18]) # column S
# Define a list (p) to contain the cumulative
# distribution values and a list (f) to
# contain the corresponding fluence values.
p=list()
f=list()
# prepare  to calculate the fit to the
# cumulative distribution at 1000
# logarithmically spaced points.
lfmin=math.log10(fmin)
lfmax=math.log10(fmax)
lf=lfmin
dlf=(lfmax-lfmin)/999
#
# calculate the fit at 1000 points
#
for i in range (1000):
# Calculate the fluence from lf.
Flu=math.pow(10.,lf)
# Use the Frechet fit above Fluence.
if Flu > Fluence:
pp=a*((Flu**(-b))-(d**(-b)))/((c**(-b))-(d**(-b)))
else:
# Use the log polynomial fit below Fluence
# and above pflu.
if Flu > pflu:
sume=0.0
for k in range(7):
sume=sume+coef[k]*math.pow(lf,6-k)
pp=math.pow(10.,sume)
else:
# Use power law fit below pflu.
pp= min(1.0,Amp*math.pow(Flu,Index))
# Store the spectrum
# Put the fluence in the list (f).
# Put the cumulative distribuion in list (p).
p.append(pp)
f.append(Flu)
# increment the log F value and end the
# For-loop.
lf=lf+dlf
# Store the fits for all the cumulative
# distributions
pch.append(p)
fch.append(f)
Scale.append(Sc)
#
#----------------------------------------------------------------
#
# Calculate the mean number of episodes expected during
# a mission
#
# Enter the year that the mission begins
#
year = float(raw_input("Enter the year your mission begins (e.g. 2015.6) "))
#
# Check to be sure the year is not before 1954
#
if year < 1954:
print 'The launch date of the mission must 1954 or later.'
print 'The launch date has been set to 1954.'
year = 1954.0
#
# Enter the mission duration
#
duration = float(raw_input("Enter the mission duration in years (e.g. 3.5) "))
#
# Check that the duration is at least two weeks.
#
if duration < 0.0385:
print 'The mission duration is shorter than a typical episode of solar energtic paricle episodes.'
print 'The mission duration has been set to 0.0385 years.'
duration = 0.0385
#
# Check to to be sure the launch date + duration is before 2054.
#
if int(year+duration) > 2052:
print 'The mission must end during the year of 2052 or earlier.'
print 'The launch date has been adjusted to end the mission in 2052.'
year=2052.999-duration
print 'The adjusted launch date is',year,'.'
#
print 'Launch Date and Mission Duration',year,' and',duration, 'years.'
#
# Use the annual sunspot number as a proxy for the episode
# frequency from 1954 through 1973. From 1974 through
# 2013 use the actual number of episodes in each year.
# From 2014 through 2020 use the predicted annual sunspot
# number. After 2021 use the fit to the 11-year solar
# cycle.
#
# Read in Mean_Annual_Episode_Frequency.csv. This file
# contains the annual episode frequencies through
# 2052 obtained as descibed above.
#
nyr=len(open(inmep,'r').readlines())
#
# Read 'inmep' as a .csv file, so that each line is a list
#
read = csv.reader(open(inmep))
#
# Create 'mepi' as a list of lists to contain the mean annual
# episode rates.
#
mepi=list()
#
# Append the lines in 'infil' to the list.
#
for line in read:
mepi.append(line)
#
# Set a reference year, ryr, for the year in which the annual
# mean epsiode rate input file starts.
#
ryr=1953.751
#
# Determine the year in which the mission starts, iy and the
# year in which it ends, jy.
#
iy=int(year-ryr)
jy=int(year+duration-ryr)
#
# If the mission begins and ends in the same year
#
if iy == jy:
mu=float(mepi[iy][1])*duration
else:
#
# If the mission ends in a later year.
#
k=jy-iy
#
# First, add the part of the calendar year in which
# the missiion began that is during the mission.
#
mu=float(mepi[iy][1])*(float(mepi[iy+1][0])-year)
#
# If ends in the second year after it begins or later,
# add the number of whole years
#
if k > 1:
for h in range (iy+1,iy+k):
mu=mu+float(mepi[h][1])
#
# Add the part of the last calendar year of the mission
# that is during the mission
#
mu=mu+float(mepi[jy][1])*(year+duration-float(mepi[jy][0]))
#
print 'Mean number of episodes expected during the mission ',mu
#
# Enter the desired confidence level
#
cl = float(raw_input("Enter the desired confidence level (e.g. 0.9 for a 90% confidence level) "))
#
#
# Now look up the fluences in each channel, taking into account that
# the mean frequency of episodes decreases with increasing channel
# number.
#
# 'flu' is defined as a list to contain the fluences in each energy
# Channel of the upper bound proton fluence spectrum.
#
flu=list()
#
# Find the upper bound fluences in each energy channel
#
for j in range (nch):
#
# Calculate the target cumulative probability, 'prob'. 
prob = -math.log(cl)/(mu*Scale[j])
# If prob < pch[j][999] then its less than the cumulative 
# probability of the largest fluence in this energy channel
if prob < pch[j][999]:
# extrapolate using the Frechet Fit.
# a through d are the Frechet parameters
a=float(lines[j][1]) #Column B
b=float(lines[j][2]) #Column C
c=float(lines[j][3]) #Column D
d=float(lines[j][4]) #Column E
flue=math.pow(((prob/a)*((c**(-b))-(d**(-b)))+(d**(-b))),(-1/b))
print 'The fluence in channel ', j+1, ' is being extrapolated.'
else:
# if prob is nearly one then the corresponding fluence is near
# background.
if prob > .99:
print 'The fluence for channel', j+1,'is near the background level and will be reported as zero.'
flue=0.0
else:
# Find the average of the nearest values of 'pch[j]' above
# and below 'prob'
im=0
for i in range (999):
if(pch[j][i] > prob):
im=i
flue=fch[j][im]+(fch[j][im+1]-fch[j][im])*(pch[j][im]-prob)/(pch[j][im]-pch[j][im+1])
flu.append(flue)
#
# Write the Upper Bound Proron Fluence Spectrum to a file
#
spec = address+'Upper_Bound_Spectrum.csv'
#
# Open the output file
#
putout = open(spec,'w')
for j in range (nch):
line = str(lines[j][0])+','+str(flu[j])+'\n'
putout.write(line)
putout.close()    
dfn.close()

\end{lstlisting}
\doublespacing

\chapter{Episodes Per Year Calculator}
\label{episodeCalc}
This appendix will provide the Python code for calculating the number of episodes per year. This program produces the .csv file that is read into the probabilistic model that contains the number of episodes per year for the 100 years covered in the model. This program reads in a .csv file that contains the monthly sunspot numbers for October 1953 to September 1974 that are found on NOAA's website \citep[see][]{sunspot}. This program also uses the sunspot proxy fit and the 11-year solar cycle fit that were detailed in \sectionref{episodes}.

\singlespacing
\begin{lstlisting}
# Program to calculate the mean number of episodes per year from October 1,
# 1953 to September 30, 2053.
# This program will use the sunspot proxy for the years 1953-1973, then real
# data for 1973-2013, the sunspot proxy for 2013-2019, and then a 11 year
# solar cycle for 2019-2053.

# Import the needed packages for this package
import math
import csv
import urllib2

# Open up the correct directory and Monthly Sunspot Numbers.csv file.
rootlst=list()
rootlst.append('C:\\Documents and Settings\\Jim\\My Documents\\Dropbox\\Probabilistic SPE Model\\CCMC\\')
rootlst.append('C:\\Users\\Jim Adams\\Dropbox\\Probabilistic SPE Model\\CCMC\\')
rootlst.append('C:\\Documents and Settings\\zach\\Desktop\\Dropbox\\Probabilistic SPE Model\\CCMC\\')
rootlst.append('D:\\Dropbox\\Probabilistic SPE Model\\CCMC\\')
yfl = 'Monthly Sunspot Numbers.csv'
address = rootlst[0]
infil = address + yfl
try:
dfn = open(infil, 'r')
except IOError:
address = rootlst[1]
infil = address +yfl
try:
dfn = open(infil, 'r')
except IOError:
address = rootlst[2]
infil = address +yfl
try:
dfn = open(infil, 'r')
except IOError:
address = rootlst[3]
infil = address +yfl
try:
dfn = open(infil, 'r')
except IOError:
print('None of these contain', yfl)
exit(0)
dfn.close()

# Find out how many full years of sunspot data we have.
monthly = list()
# Define a function to read in the input file.
read = csv.reader(open(infil))
# Now place each month's sunspot total into the list labeled monthly.
for line in read:
monthly.append(line)
# Calculate the number of years by taking the length of the monthly list and
# dividing by 12. Round down to the nearest integer.
sunspotYears = int(len(monthly)/12)
print 'Numbers of years: ', sunspotYears

# Next, we will calculate the number of sunspots observed during each year.
# First, create a list to put the data into.
sunspots = list()
# Since the monthly sunspot list starts in 1953, the first value in the sunspot
# list will correspond to the number of sunspots in 1953, the second value will
# be for 1954, etc. up until the present.  This will not count up the sunspot
# number for the current year.  In this code, the year 1953 refers to the time
# period October 1, 1953 to September 30, 1954.
for i in range(sunspotYears):
# Create a term to add the sunspots in.
sunspot = float(0)
# Now for each year, add up the 12 month's values to get a total
# number of sunspots in the year.
for j in range(12):
sunspot = sunspot + float(monthly[12*i+j][1])
# Finally, put the yearly total into the sunspots list.
sunspots.append(sunspot)

# Now, we have to go online to get the sunspot predictions up through 2020.
# Define the website to go to find the data.
downloaded_data  = urllib2.urlopen('http://www.ngdc.noaa.gov/stp/space-weather/solar-data/solar-indices/sunspot-numbers/predicted/table_international-sunspot-numbers_monthly-predicted.txt')
# Create two list that will be used to manipulate the data.
web = list()
monthlyPredictions = list()
# We will download and manipulate one line at a time.
for line in downloaded_data.readlines():
# Save the line under a different name.
lnes = line
# Parse the line everywhere there is a space.
lnes = lnes.split(' ')
# Create a new list for the data we want to keep from our parsed line.
lfn = list()
for n in range(len(lnes)):
# We only want to keep the sections of data that have a value in it.
if lnes[n] != '':
# Record these good values in our newest list, lfn.
lfn.append(lnes[n])
# Finally, we can record each line into our list called web.
web.append(lfn)
#print(web)

# Now we get just the pieces of data that we want, the monthly predictions.
# We record these lines into our second list, called monthlyPredictions.
for i in range((len(web)-2)/3):
monthlyPredictions.append(web[3*i+2])
#print(monthlyPredictions)

# This gets rid of the carrage-return-linefeed at the end of each line.
for i in range(len(monthlyPredictions)):
j = len(monthlyPredictions[i])
monthlyPredictions[i][j-1]=monthlyPredictions[i][j-1][:-2]
#print(monthlyPredictions)

# This section gets rid of the weird symbol out in front of the any year.
for i in range(len(monthlyPredictions)):
if monthlyPredictions[i][0] == '\t':
ex = list()
for j in range(1,14):
ex.append(monthlyPredictions[i][j])
monthlyPredictions[i] = ex
#print(monthlyPredictions)

# This section turns the strings into numbers. 
for i in range(len(monthlyPredictions)):
for j in range(len(monthlyPredictions[i])-1):
monthlyPredictions[i][j+1] = float(monthlyPredictions[i][j+1])
#print(monthlyPredictions)


# Define the function to use our sunspot proxy code.  
def SunspotProxy(slope, intercept, Q, sunspots, episodes):
# Turn the variables into numbers using the float command.
slope = float(slope)
intercept = float(intercept)
Q = float(Q)
sunspots = float(sunspots)
# Now use the sunspot proxy equation K=(AN+B)exp(-Q(AN+B)) where
# A is the slope of a straight line, B is the y-intercept, N is the
# annual sunspot number, K is the mean number of episodes per year,
# and Q is our deadtime coefficient.
epi = (slope*sunspots+intercept)*math.exp(-Q*(slope*sunspots+intercept))
# Finally, add the calculated number of episodes in this year to the list
# named episodes.
episodes.append(epi)

# These are our calculated parameters for the sunspot proxy for a year starting
# on October 1 and going to September 30.  These parameters can be updated in
# the future when more data becomes available.
slope = 0.0491
intercept = -1.6122
Q = 0.0151

# Now we have to deal with the 11 year solar cycle section of the prediction.
# For this, we will use the code that Jim Adams wrote from the program named
# 'Episodes per Year.py' found in Dropbox\Probabilistic SPE Model\CCMC folder
# on Dropbox.  It will be tweaked a little to work properly for this program.

# Set a reference year, ryr, for the solar cycle to come closest
# to matching the episode frequency minima in the data, assuming
# 11 year cycles. ryr is the the date of the beginning of the first
# year of the next solar cycle.
#
ryr=2019.751
#
# Calculate the average number of episodes in each calendar year from
# 2019 to 2053. First data-in the mean number of episodes in 
# each year of the 11-year solar cycle that best fits the 
# data on the frequiency of episodes. The next solar cycle starts will start
# October 1, 2019.  The average number of episodes in each 
# solar cycle year # is in 'Episodes-per_year' list. The first 
# entry in this list is for the first year following solar 
# minimum.
#
Episodes_per_year=[5.25,9.75,17.00,22.75,25.00,19.00,26.50,20.60,14.67,7.33,8.17]
#
# Create a list to hold each year, yr
yr=list()
# Create a list to hold the mean number of episodes in each year, mepisodes
mepisodes=list()
# There are 34 years from 2019 to 2053. This is the span of years for which
# the mean number of episodes per year will be calculated.
years=34
# Duration in all cases is 1 year
duration=1.0
#
for i in range (years):
year0=ryr+float(i)
year=year0
#
# index is the years between 2019.751 and the launch date
# truncated to an integer.
#
index=int(year-ryr)
#
# end is the time when the mission ends
#
end=year+duration
#
# Initialize the mean number of episodes expected during the 
# mission, mu, to zero. Initialize flag to 1. Flag is set 
# to zero to escape the 'While' loop. ssyr is the fraction 
# of each solar cycle year during the mission. Initialize 
# ssyr to zero and initialize y0 to zero.
#
mu=0.0
flag=1
ssyr=0.0
y0=0.0
#
# Use a While-loop to determine the mean number of episodes
# during the mission recursively
#
while flag > 0:
# Find the solar cycle year in which the mission atarts 
# and the fraction of that year covered by the mission.
ssyr=ryr+index+1-year
# Find the length of the remainder of the mission after the
# end of the solar cycle year in which the mission began.
z=end-ryr-index-1
# There are three cases, depending on the value of Z
if z < 0: # The mission ends in the solar cycle year where
# 'year' occurs.
ssyr=ssyr+z # correct ssyr by removing the remainder of
# the solar cycle year not included in the mission.
flag=-1 # Set flag to escape the while loop.
else: # There are two cases if the mission extends into the
# next solar cycle year.
if z < 1-ssyr: # The duration of the remainder of
# the mission is less than one year.
y=z # Store the part of the mission that extends into
# the next solar cycle year in y.
flag=0 # Set the flag to add to mu the mean number
# of epsiodes occuring in the cycle year containing
# the remaining portion of the mission and exit the
# while loop.
else: # The remaining duration of the mission after 'year'
# is > one year long. Store the part of the mission
# between 'year' and 'year'+1 that occurs during the
# solar cycle year following the one in which 'year'
# occurs in y.
y=1-ssyr
# Add to mu the mean number of epsiodes for the portion of
# the solar cycle year in which 'year' occurs that was
# during the mission.
mu=mu+(ssyr+y0)*Episodes_per_year[index%11]
if flag > -1:
index=index+1 # Set index for the next solar cycle year.
year=year+1 # Begin with the second year of the mission.
y0=y # Transfer the stored time from y to y0.
if flag == 0: # Add to mu the mean number of episodes
# occuring during the portion of the mission that
# extends into the next solar cycle year after the one
# in progress in 'year'.
mu=mu+y*Episodes_per_year[index%11]
yr.append(year0)
mepisodes.append(mu)

# Now we have coded all the tools we need to create a list of the number of
# episodes per year.  We will create a new list to store these values.
episodes = list()

# First we will use the sunspot proxy to calculate the episodes per year
# for the years 1953 to 1972.
for i in range(20):
SunspotProxy(slope, intercept, Q, sunspots[i], episodes)

# For 1973, we need to approach this year slightly different than the sunspot
# proxy.  We have data from GME starting on November 1, 1973. Instead of using
# the sunspot proxy for the enitre year, we will use it for one month that
# we are missing.  First create a list to store this year's sunspot proxy
# calculation.
epi1973 = list()
# Now do the sunspot proxy calculations.
SunspotProxy(slope, intercept, Q , sunspots[20], epi1973)
epis1973 = float(epi1973[0])/12.0
# epis1973 now holds the value for one month worth of events.  Now take this
# value and add the number of episodes from the other 11 months that were
# observed using GME.  There were 17 episodes observed in the rest of the year.
episodes1973 = epis1973 + 17.0
# So episodes1973 now contains how many episodes there were in the year 1973.
# Now add this number to the full list of episodes for each year.
episodes.append(episodes1973)
# Now we want to add in the number of observed episodes from 1974 to 2012.
# The list ObservedEpisodes will store the observed episodes for these years.
# These values are obtained from the 'Solar Cycle.xlsx' file in the CCMC 
# folder on Dropbox.
ObservedEpisodes=[10,9,11,27,24,21,31,26,16,14,9,6,6,21,22,21,20,25,16,9,7,2,5]
ObservedEpisodes=ObservedEpisodes+[24,25,24,29,21,21,17,19,1,4,0,0,11,18,26,14]
# Now we can add these to the episodes list.
for j in range(len(ObservedEpisodes)):
episodes.append(ObservedEpisodes[j])
# Just to recap, the list 'episodes' now contains the number of episodes per
# year from October 1, 1953 to September 30, 2013.

# Since there is not enough data to complete the next year, we will now use
# the sunspot proxy and the predicted sunspot numbers to calculate the number
# of episodes per year for the years 2013-2018.
# First, we have to calculate the current year.  If we add 'sunspotYears'
# and the length of 'ObservedEpisodes' to 1953, it will tell us the
# current year.
currentYear = 1953 + len(ObservedEpisodes) + sunspotYears
# This gives the year 2013, which is the current year.
# Now we have to determine how many years we need to do this method of
# of calculations for.
predYears = 2019 - currentYear
# Now determine the annual sunspot numbers for these predicted years.
for n in range(int(predYears)):
# This chooses each year that a prediction is needed.
for s in range(len(monthlyPredictions)):
# This scans each line of 'monthlyPredictions' to find the predictions
# for a given year that matches the year we are looking for.
if monthlyPredictions[s][0] == str(currentYear + n):
# When the program matches up the years, it will enter this for
# loop.  This loop will add up all the sunspots predicted in a
# given year in the future.
# REMEMBER: The year still goes from October 1 to September 30 of
# the following year.
# Start by adding up October to December predictions.
annualSunspot=monthlyPredictions[s][10]+monthlyPredictions[s][11]
annualSunspot = annualSunspot + monthlyPredictions[s][12]
# Now continuing adding up the total predictions for the rest of
# the year, January through September. (These values are in the
# next year)
for k in range(1,10):
# The s+1 in the following line causes the January to September
# values to comes from the next calendar year.
annualSunspot = annualSunspot + monthlyPredictions[s+1][k]
# Finally, use the sunspot proxy to calcualte the episodes per year
# for these years.
SunspotProxy(slope, intercept, Q, annualSunspot, episodes)

# The list episodes contains the episodes per year from October 1, 1973 to
# September 30, 2019.  We have already calculated the episodes per year from
# October 1, 2019 to Spetember 30, 2053 above.  These values are stored in the
# list mepisodes.  So all we have to do is add this list onto the episodes list.
for t in range(len(mepisodes)):
episodes.append(float(mepisodes[t]))

# We have finally finished the list of 100 years worth of episodes.
# The last step is to print it out into a csv file that can be saved and used
# in the future.
# To do this, we first have to create an output file.
out = address + 'Mean_Annual_Episode_Frequency.csv'
# Open the output file.
putout = open(out, 'w')
# Now we need to form each line and then write it out to the output file.
# Since there are 100 years worth of data, we will have to write out 100 lines.
for d in range(100):
# Each line has 2 columns, the first being the year while the second
# contains the calculated number of episodes for that year.
line = str(1953.751 + d) + ',' + str(episodes[d]) + '\n'
# Write out the line to the output file.
putout.write(line)
# Close the output file.
putout.close()

\end{lstlisting}
\doublespacing


\end{appendices}
\backmatter

\begin{singlespace}
	\bibliography{Robinson}
\end{singlespace}

\end{document}